\documentclass{aa}
\usepackage{graphicx}
\usepackage{txfonts}

\usepackage{color}
\usepackage{array}
\usepackage{caption}
\usepackage{subcaption}
\usepackage{amsmath,amssymb,amsfonts}
\usepackage{gensymb, footnote}
\usepackage{sidecap}
\usepackage{changepage}
\usepackage{tabu}
\usepackage[toc,page]{appendix}
\usepackage{multirow}
\usepackage[hidelinks]{hyperref}

\usepackage{tablefootnote}
\usepackage{threeparttable}
\usepackage{upgreek}

\title{BHs in the Milky Way with Gaia}
\author{Soetkin Janssens}
\date{March 2021}

\newcommand{\Teff}{T_{\text{eff}}}

\newcommand{\Modot}{M_{\odot}}

\newcommand{\Rodot}{R_{\odot}}

\newcommand{\LC}{_{\text{LC}}}
\newcommand{\BH}{_{\text{BH}}}

\newcommand{\kms}{km\,s$^{-1}$}

\newcommand{\Tabref}{Table\,\ref}
\newcommand{\Figref}{Fig.\,\ref}
\newcommand{\Secref}{Sect.\,\ref}
\newcommand{\Eqref}{Eq.\,\eqref}

\newcommand{\Figrefs}{Figs. \ref}

\newcommand{\nextline}{\\\indent}

\newcommand{\Gaia}{\textit{Gaia}}
\newcommand{\Gaias}{\textit{Gaia}\space}

\begin{document}
   \title{Uncovering astrometric black hole binaries with massive main-sequence companions with Gaia} \titlerunning{Astrometric black hole binaries with Gaia}

   \author{S. Janssens \inst{1} 
           \and T. Shenar\inst{1}
           \and H. Sana\inst{1}
           \and S. Faigler \inst{2}
           \and N. Langer\inst{3,4}
           \and P. Marchant\inst{1}
           \and T. Mazeh \inst{2}
           \and C. Sch\"urmann\inst{3,4}
           \and S. Shahaf\inst{5}
          }

   \institute{Institute of Astronomy, KU Leuven, Celestijnenlaan 200D, 3001 Leuven, Belgium\\ \email{soetkin.janssens@kuleuven.be}
            \and
         School of Physics and Astronomy, Tel Aviv University, Tel Aviv 6997801, Israel
         \and
         Argelander-Institut f\"ur Astronomie, Universit\"at Bonn, Auf dem H\"ugel 71, 53121 Bonn, Germany
         \and 
         Max-Planck-Institut f\"ur Radioastronomie, Auf dem H\"ugel 69, 53121 Bonn, Germany
         \and
         Department of Particle Physics and Astrophysics, Weizmann Institute of Science, Rehovot 7610001, Israel
         }    

   \date{}

 
  \abstract
   {In the era of gravitational wave astrophysics and precise astrometry of billions of stellar sources, the hunt for compact objects is more alive than ever. Rarely seen massive binaries with a compact object are a crucial phase in the evolution towards compact object mergers. With the upcoming \Gaias data release (DR3), the first \Gaias astrometric orbital solutions for binary sources will become available, potentially revealing many such binaries.
   }
   {We investigate how many black holes (BH) with massive main-sequence dwarf companions (OB+BH binaries) are expected to be detected as binaries in \Gaias DR3 and at the end of the nominal 5-yr mission. We estimate how many of those are identifiable as OB+BH binaries and we discuss the distributions of the masses of both components as well as of their orbital periods. We also explore how different BH-formation scenarios affect these distributions.}
   {We apply observational constraints to tailored models for the massive star population, which assume a direct collapse and no kick upon BH formation, to estimate the fraction of OB+BH systems that will be detected as binaries by \Gaia, and consider these the fiducial results. These OB+BH systems follow a distance distribution according to that of the second Alma Luminous Star catalogue (ALS II). We use a method based on astrometric data to identify binaries with a compact object and investigate how many of the systems detected as binaries are identifiable as OB+BH binaries. Different scenarios for BH natal kicks and supernova mechanisms are explored and compared to the fiducial results.}
   {In the fiducial case we conservatively estimate that 77\% of the OB+BH binaries in the ALS II catalogue will be detected as binaries in DR3, of which 89\% will be unambiguously identifiable as OB+BH binaries. At the end of the nominal 5-yr mission, the detected fraction will increase to 85\%, of which 82\% will be identifiable. The 99\% confidence intervals on these fractions are of the order of a few percent. These fractions become smaller for different BH-formation scenarios.}
   {Assuming direct collapse and no natal kick, we expect to find around 190 OB+BH binaries with \Gaias in DR3 among the sources in ALS II, which increases the known sample of OB+BH binaries by more than a factor of 20, covering an uncharted parameter space of long-period binaries ($10 \lesssim P \lesssim 1000\,$d). Our results further show that the size and properties of the OB+BH population that is identifiable using \Gaias DR3 will contain crucial observational constraints to improve our understanding of BH formation. An additional $\sim$5 OB+BH binaries could be identified at the end of the nominal 5-yr mission, which are expected to have either very short ($P \lesssim 10\,$d) or long periods ($P \gtrsim 1000\,$d). 
   }


   \maketitle

\section{Introduction}
Massive stars ($M_{\text{initial}} \gtrsim 8\Modot$) play an important role in the evolution of the Universe and its galaxies. Their powerful radiation fields ionise their surroundings and their strong winds enrich the material in their neighbourhood, affecting star formation rates and chemical compositions of their host galaxies \citep[e.g.][]{Heckman_1990, Bresolin_2008, Hopkins_2014}. They typically end their lives with a powerful supernova, leaving behind a compact object remnant -- a black hole (BH) or a neutron star (NS). Recently, the detection of compact object mergers \citep[e.g.][]{Abbott_2016b, Abbott_2016a, Abbott_2017} have reignited the interest in the evolution of massive stars and the formation of their remnants, as massive binaries with a BH component are a crucial phase in the evolution towards compact object mergers. \nextline
Several independent studies \citep[e.g.][]{Shapiro_1983, vand_den_Heuvel_1992, Brown_1994, Timmes_1996, Samland} estimate that $\sim$10$^8$ -- $10^9$ stellar-mass BHs reside in our Galaxy. One way of finding these BHs is through binaries. Most massive stars are found in binaries or higher order multiple systems and will interact with their companion during their evolution \citep{Sana_2012, Sana_2014, Moe_2017}. \citet{Langer_2020} predicted that $\sim$3\% of these massive O- and early B-type stars in binaries have a BH companion, resulting in an estimated $\sim$1200 OB+BH systems in the Milky Way. Despite the enormous amount of predicted BHs, only 59 candidate BHs have been reported so far, all of which in interacting binaries as X-ray sources \citep{Corral-Santana_2016} and most of them with low-mass luminous companions. \nextline
Finding BHs in non-interacting binaries appears to be notoriously difficult. Two very recent examples are the BHs claimed to be present in LB1 and HR6819 \citep[][respectively]{Liu_2019, Rivinius_2020}. Soon after their publications, several research teams argued that these two systems did not contain BHs, but rather that these are binaries containing a stripped star and a Be star or that the unseen companion is a close binary \citep[e.g.][]{Abdul-Masih_2020, Bodensteiner_2020, El_badry_2020_LB1, Irrgang_2020, Mazeh_2020, Shenar_2020}. However, there is still an ongoing debate about the nature of these systems \citep{Liu_2020, Lennon_2021}. Very few other quiescent BH candidates exist in the Milky Way. Two are reported in \citet{Casares_2014} and \citet{Giesers_2018}: a 10 -- 16\,$\Modot$ Be star with an orbital period of 60\,days, and an 0.81\,$\Modot$ star with an orbital period of 167\,days, respectively. Two more systems with low-mass giant luminous companions and periods below 100\,days are reported in \citet{Thompson_2019} and \citet{Jasinghe_2021}.
\nextline
The lack of BH candidates and the difficulty of showing that a BH indeed resides in those candidates \citep[e.g.][]{Trimble_1969,Liu_2020, Lennon_2021} indicate that other methods are needed to detect these systems. The European Space Agency mission \textit{Gaia} \citep{Gaia_colab_2016_mission}, which is an astrometric all-sky survey, provides one such very promising method for the detection of OB+BH systems. This was already mentioned by, for example, \citet{Gould_2002} when they gave predictions on BH detections for \Gaia's predecessor \textit{Hipparcos}. At the end of its nominal 5-yr mission, \Gaias will provide astrometric parameters for $\sim$ 1 billion sources brighter than $G$ magnitude $\sim$20 with a precision of the order of $10$--$100$\,microarcseconds or $\mu$as. By surveying an unprecedented 1\% of the stellar population in the Milky Way, \textit{Gaia} will bring a new era in the detection and analysis of the population of OB+BH binaries.\nextline
The third and most recent data release of \textit{Gaia} is split into two parts: an early (EDR3) and a full data release (DR3), based on 34 months of observations \citep[][]{Gaia_collaboration_2020_summary}. Currently, only EDR3 is published, which provided updated astrometry and photometry of $\sim$1.8 billion sources with a precision of the order of $0.1$--$1$ mas \citep{Lindegren_2020}. With its expected release date in the second quarter of 2022, DR3 will provide us with the first \Gaias astrometric solutions of binary systems.\nextline
Several studies \citep[e.g.][]{Breivik_2017,Breivik_2019,Mashian_Loeb_2017, Yalinewich_2018, Yamaguchi_2018, Wiktorowicz_2019} estimated how many binaries with a BH component \textit{Gaia} will detect by the end of its nominal 5-yr mission. The numbers range from hundreds to ten thousands of systems. Identifying so few BH systems in data of more than one billion sources is reminiscent of finding a few needles in a haystack. These predictions are based on the estimated precision of the data at the end of the nominal 5-yr mission. However, the precision that is reached in DR3 allows for a similar analysis. With the goal of constraining massive-star evolution, in this work we focus on the massive star population, i.e., potential progenitors of BH+BH/NS. \nextline 
There are multiple reasons why we could not simply scale the results obtained by previous studies. Firstly, their simulations were performed with the inclusion of initial stellar masses below 8\,$\Modot$, whereas we only focus on the stellar population with $M_{\text{initial}} \gtrsim 8\Modot$. Secondly, the simulations of \citet{Breivik_2017} and \citet{Wiktorowicz_2019} did not only consider binaries with main-sequence dwarf luminous companions (MS+BH binaries), but also those with giant luminous companions, and \citet{Breivik_2019} considered only the latter. Thirdly, the simulations of some previous works 
did not include important massive star physics. For example, \citet{Mashian_Loeb_2017} do not take into account orbital angular momentum change due to mass transfer between the components. \citet{Yamaguchi_2018} neglect wind mass loss from both stars. Binary interactions and wind mass loss are important aspects of the evolution of massive stars, the latter especially in the late stages of their evolution. Hence, these earlier simulations are not adapted to our purpose.
\\
\nextline 
The goal of this paper is threefold. First, we estimate the fraction and absolute number of OB+BH systems in the second Alma Luminous Star Catalogue \citep[ALS II;][]{Gonzalez_2021} that will be detected as astrometric binaries in \Gaias DR3 and in the data release at the end of the nominal 5-yr mission, DR4. The ALS II is an established catalogue of massive OB-type stars in the \Gaias catalogue. While it is not complete and suffers from non-trivial selection effects (see \Secref{sec_distances}), it is the most comprehensive catalogue of massive OB stars existing to date, with over 13\,000 entries. In our simulations, we consider a direct collapse -- all baryonic mass ends up contributing to the final BH mass -- and no natal kick to be the fiducial BH-formation model. Furthermore, the predictions are based on the mass-magnitude relation for main-sequence dwarf OB stars (see \Secref{sec_mass-magnitude_relation}). Secondly, we investigate how many of the detectable systems can be unambiguously identified as OB+BH systems. Finally, we look at the expected observable distributions of the masses of both components and the orbital period of the binary systems, while also exploring the effect of different BH formation scenarios on these distributions.\nextline
Section \ref{sec_simulations_title} presents the binary population synthesis simulations on which we rely. Section \ref{sec_Gaia_cuts_title} shows the distribution of the simulated OB+BH systems according to the ALS II and explains the observational constraints imposed on their detection. We explain how we will identify the \Gaias OB+BH systems in \Secref{sec_shahaf_title}. The detection and identification fractions are presented in \Secref{sec_BH_distributions_title}, together with the expected distributions of the masses and the period. We discuss our results and their uncertainties in \Secref{sec_discussion} and end with a summary in \Secref{sec_summary}.

\begin{figure*}
    \centering
    \includegraphics[width = \textwidth]{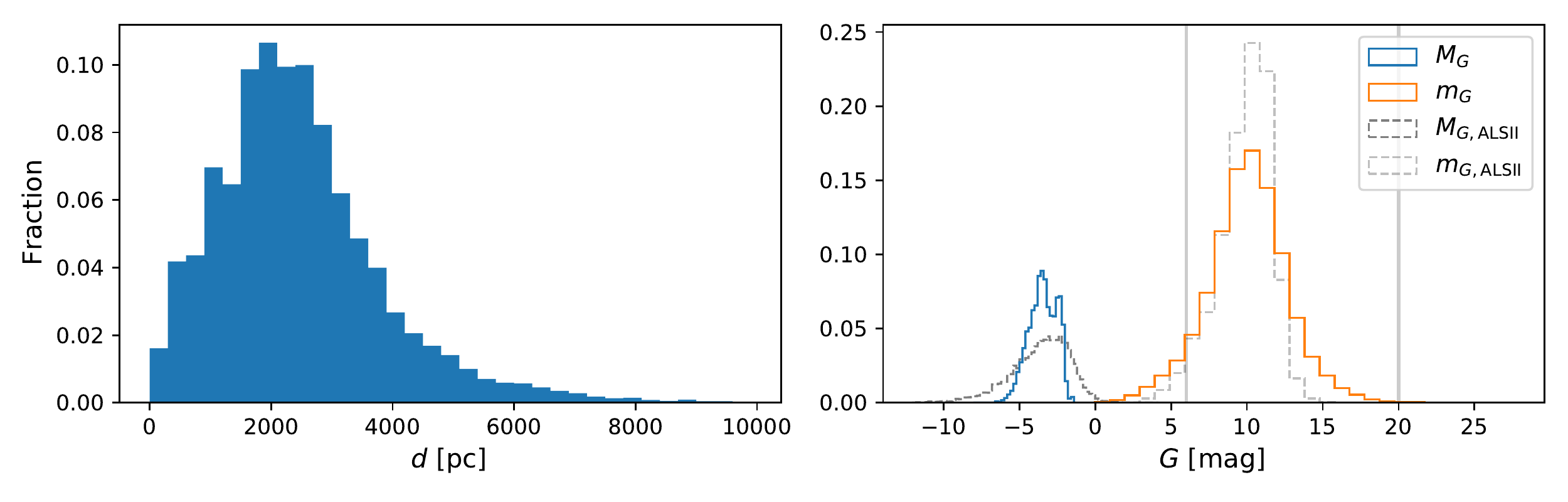}
    \caption{Left: Distances of the massive OB stars in the ALS II catalogue. Right: Absolute (open blue) and apparent (open orange) magnitudes of the full sample of simulated OB+BH systems, as well as those of the massive stars in the ALS II catalogue (black and gray dashed, respectively). }
    \label{fig_distances_magnitudes}
\end{figure*}

\section{Binary evolution models} \label{sec_simulations_title}
\subsection{General framework}
The present work relies on the computations of \citet{Langer_2020}, which performed binary simulations with the stellar evolution code MESA \citep[Modules for Experiments in Stellar Astrophysics;][]{Paxton_2011}. Their simulations are tailored for massive stars and use the observed distribution of initial conditions for massive stars as determined by \citet{Sana_2012}:
\begin{equation}
    \begin{split}
        \text{Prob}&(\log P_{\text{i}}) \propto (\log P_{\text{i}})^{-0.55},\\
        \text{Prob}&(q_{\text{i}}) \propto q_{\text{i}}^{-0.1}.
    \end{split}
\end{equation}
Here, $P_{\text{i}}$ is the initial period and $q_{\text{i}}$ is the initial mass ratio, which is the initial mass of the secondary (least massive) divided by that of the primary (most massive) star. It is the primary star that will evolve into the BH. The mass of the primary follows the \citet{Salpeter_1995} initial mass function, ranging from 10 to 40 $\Modot$, where the upper mass limit comes from the assumption that more massive stars undergo large inflation during their main-sequence evolution \citep[][]{Brott_2011}, inducing a common-envelope phase that leads to the merger of the binary components before the evolution towards an OB+BH system. Initial mass ratios range from 0.25 to 0.975 and orbital periods from 1.41 to 3160 days, focusing on the OB+BH population originating from post-mass-transfer binaries. The simulated systems with the longest periods evolve into a common-envelope phase, where it is assumed a merger follows before the core collapse of the primary.\nextline
For main-sequence stars that have not yet interacted, the most massive star is also the most luminous. Hence, the above-stated way of defining the mass ratio is equal to the photometric mass ratio, which is the mass of the least luminous companion (secondary) divided by that of the most luminous (primary). For the remainder of this work, we work with the photometric mass ratio, primary, and secondary. \nextline
Most of the OB+BH progenitors undergo a phase of mass transfer, after which the donor star (the BH progenitor) becomes a stripped star. Given the mass range of BH progenitors, the stripped star is expected to be a Wolf-Rayet (WR) star — a He-burning star exhibiting powerful stellar winds \citep{Shenar_2020_WR}. 
Eventually, a massive enough WR star will collapse into a BH, leaving behind an OB+BH system. In their models, \citet{Langer_2020} assume stars with a He-core mass $ M_{\text{Hec}}>6.6\Modot$ at core carbon exhaustion become BHs through a direct collapse, meaning that all baryonic mass ends up contributing to the BH and no mass is lost through neutrinos either. Moreover, no natal kick is assumed. Stars with $ M_{\text{Hec}}\leq6.6\Modot$ are not considered BH progenitors.
\nextline
The simulations of \citet{Langer_2020} were, however, performed with a metallicity representative of the Large Magellanic Cloud (LMC), i.e. about half solar. If we want to obtain the same set of systems in the Milky Way, one of the major differences would be the increase of mass lost through winds, as the winds of massive stars become stronger in higher metallicity environments \citep[e.g.][]{Vink_2001,Crowther_2006,Mokiem_2007}. One of the downsides in using detailed model evolution codes, such as MESA, is that it requires a large amount of time and computing power to run a grid of models, such as those of \citet{Langer_2020}, and that uncertain parameters, such as the BH-formation mechanism, cannot easily be varied. Therefore, we worked with their original simulations for the LMC and applied a correction to the final OB+BH systems to account for first-order effects due to the transition to solar metallicity. We explore the impact of different BH-mechanisms in \Secref{sec_other_scenarios}. 

\subsection{Corrections for the Milky Way}\label{sec_sims_LMC_to_MW}
We performed two adjustments for the transformation of the systems. First, for a given stellar mass, we accounted for a temperature decrease of 1000\,K \citep{Sabin-Sanjulian_2017} for the luminous companion (LC), hence also decreasing its luminosity through Stefan-Boltzmann's law
. The radius of the LC was unaltered. While some change in the radius is expected at different metallicity, the impact of this change is negligible. Moreover, the change in luminosity from the effective temperature is small and does not significantly alter the results presented in this work.\\\nextline
Second, we accounted for metallicity-dependent wind strength. Due to stronger winds in higher-metallicity environments, such as the Milky Way, the final masses of the BHs and the LCs will be lower than those in low-metallicity environments, such as the LMC. The extra mass loss also leads to an increase in period.\nextline
We used a power-law dependence of mass loss and metallicity $\dot{M} \propto Z^{0.83}$, as determined by \citet{Mokiem_2007}. In general, the wind mass loss of WR stars obeys a slightly steeper metallicity dependence \citep[e.g.][]{Hainich_2015}, which also varies for different hydrogen surface abundances. For simplicity and to neglect scatter on the metallicity dependence, we used $\dot{M} \propto Z^{0.83}$ also for the WR stars.\nextline
Since the difference in mass-loss rates of main-sequence stars is negligible, we only applied the mass-loss correction to the systems when mass transfer has fully stopped when the components detach. We used $\log R /\log R_{\text{RL}} < 0.99$ as the criterion for detachment, where $R$ is the radius of the primary star (the BH progenitor) and $R_{\text{RL}}$ is its Roche-lobe radius.\nextline 
The wind mass-loss rate for the LMC was determined through $\dot{M}_{\text{LMC}} = (M_{\text{rl-end}} - M_{\text{f}})/\Delta\tau_{\text{f,rl-end}}$, where $\Delta\tau_{\text{f,rl-end}}$ is the time difference between the SN of the primary and detachment. Through the mass-loss-metallicity relation, the mass-loss rate in the Milky Way is related to that of the LMC as
\begin{equation}
\dot{M}_{\text{MW}} = \dot{M}_{\text{LMC}}(Z_{\text{MW}}/Z_{\text{LMC}})^{0.83},
\end{equation}
with $Z_{\text{MW}}/Z_{\text{LMC}} = 2$. The final mass of a star in the Milky Way is then given by
\begin{equation}
    M_{\text{f,MW}} = M_{\text{rl-end}} - \dot{M}_{\text{MW}} \Delta\tau_{\text{f,rl-end}}.
\end{equation}
This correction was done for both components, where the final mass of the LC is its mass at primary BH formation.\\
\nextline
The final period of the system due to mass loss of both components is obtained from (see Appendix \ref{app_period_change}) 
\begin{equation}\label{eq_final_period}
    P_f = P_i\left( \frac{M_{1,i} + M_{2,i}}{M_{1,f} + M_{2,f}} \right)^2,
\end{equation}
where the subscript `$i$' indicates the initial conditions, in this case the conditions at detachment, and the subscript `$f$' indicates the final conditions of the system or the conditions right before BH formation. There are few systems which do not undergo a phase of mass transfer due to their long periods. For those systems, the initial conditions are those at the beginning of the simulations. \nextline
For the original simulations (no mass or period corrections), systems which undergo mass transfer already show a slight difference between the final period in the simulation and the period obtained with \Eqref{eq_final_period}. One of the assumptions made when using \Eqref{eq_final_period} is that both stars reside well within their Roche lobe. At detachment 
this assumption is not valid and hence \Eqref{eq_final_period} might be too simplistic. Especially for short-period systems and stars close to filling their Roche-lobe, tides affect the period change. Assuming tidal effects are metallicity independent, we apply a relative period change -- the ratio of the period in the original system $P_{\mathbf{sim,LMC}}$ and its period calculated with \Eqref{eq_final_period} $P_{\mathbf{eq,LMC}}$ -- to the final period derived with the masses in the Milky Way, such that the final period for the Milky Way systems obtained with \Eqref{eq_final_period} is multiplied with a factor $P_{\mathbf{sim,LMC}}/P_{\mathbf{eq,LMC}}$.\nextline
The applied wind-mass loss corrections do not lead to major changes in the distributions of the LC mass, the mass-ratio, and the period post-SN, between LMC and Milky-Way systems, whereas the most massive BHs have slightly shifted towards lower masses. Since $\dot{M}$ increases with increasing $M$, the amplification of $\dot{M}$ causes the most massive BH progenitors to be over-proportionally lighter. The different distributions are shown in Appendix \ref{app_LMCvsMW}.

\begin{figure}
    \centering
    \includegraphics[width = 0.45\textwidth]{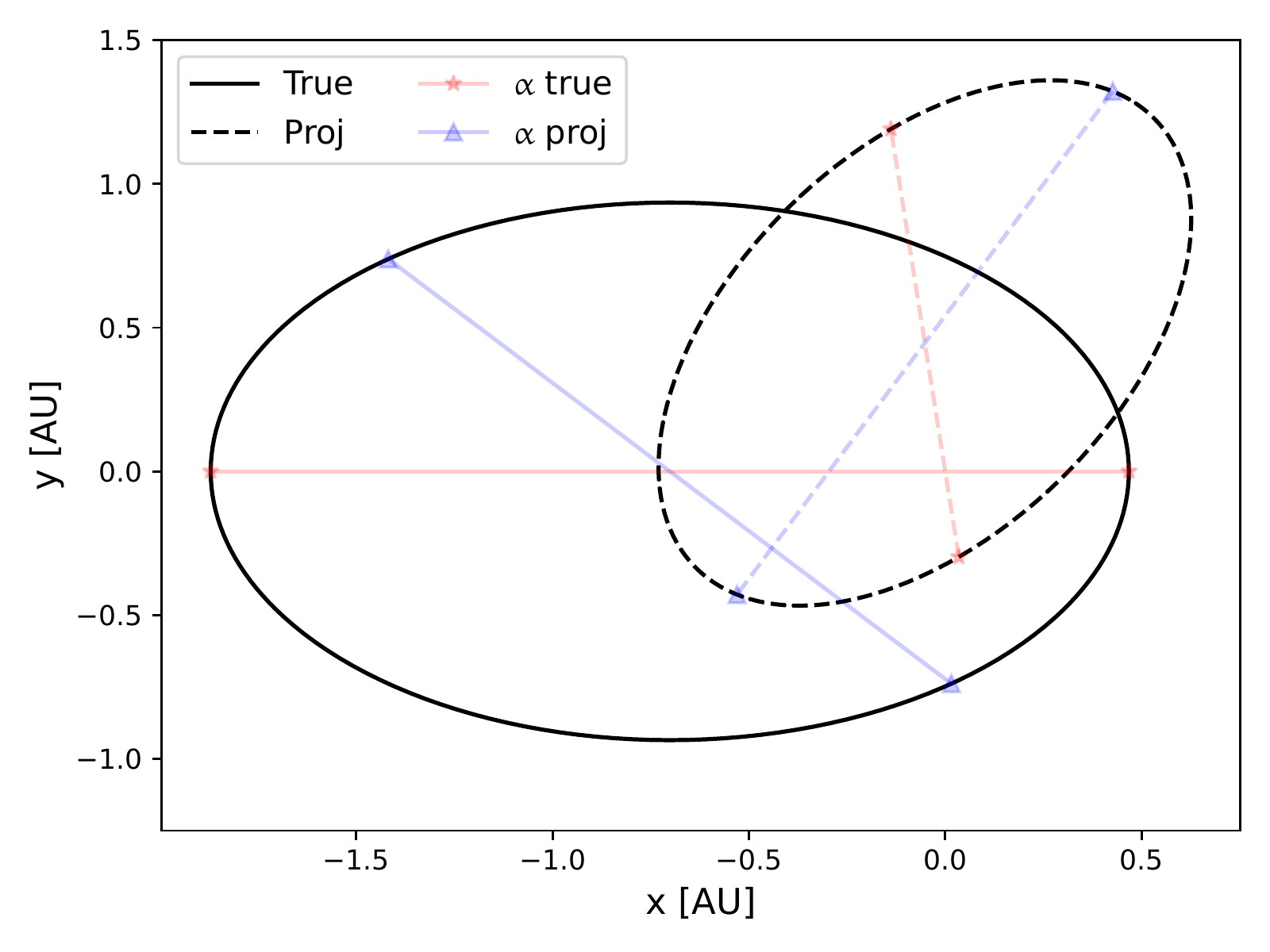}
    \caption{Orbital projection for an example of a MS+BH system with $M_{\text{BH}} = 10\Modot$, $M_{\text{LC}} = 15\Modot$, $a_{\text{LC}} = 1.17$\,au, and $e = 0.6$. The solid ellipse represents the true orbit, with the semi-major axis shown as the two stars connected by a red line. The dashed ellipse shows the orbit projected with $i = 60\degree$, $\Omega = 52.9\degree$, and $\omega = 242.4\degree$. The projected semi-major axis is shown as a red dashed line. The blue triangles connected with a dashed line show the semi-major axis of the projected orbit $\upalpha$.}
    \label{fig_projected_semimajor-axis}
\end{figure}
\section{The detectability of OB+BH systems with Gaia}\label{sec_Gaia_cuts_title}
We wish to estimate how many of the OB+BH systems that are observed by \Gaias will also be detected as binaries in DR3 and in DR4. For this, we apply several observational constraints to the simulated OB+BH systems. Observational constraints are determined by the apparent $G$-band magnitude, which depends on the distance to the system, the period of the system, and the observed astrometric signal. A discussion on the observational constraints is given at the end of \Secref{sec_estimated_numbers}.

\subsection{The distance distribution of the systems}\label{sec_distances}
The distance distribution of the simulated systems was based on the distances of known massive stars, and hence does not represent the intrinsic distribution in the Milky Way, but an observational one. A well known massive star catalogue is the Alma Luminous Star \citep[ALS;][]{Reed_2005} catalogue. \citet{Gonzalez_2021} have created the ALS\,II, in which they obtained distances for the stars in the ALS based on \Gaias DR2 data and a prior tailored for massive stars. We only used the distances of sources (the number of sources is only used later in \Secref{sec_estimated_numbers}) which have high-quality DR2 data and excluded those flagged as non-massive stars. An observational constraint on the distance is embedded in the magnitude limit. The distance distribution is shown in the left panel of \Figref{fig_distances_magnitudes}.\nextline
We note that the ALS II is not necessarily complete, but rather represents a clean sample of massive OB stars formed by compiling previous catalogues and literature sources of massive OB stars. However, for the purpose of this research, we do not require a complete sample, but instead a large and clean enough sample of known massive stars. Also the possible lack of young massive stars \citep{Holgado_2020} is not a real concern in this research, as most of the BHs are expected to have companions older than $\sim$5\,Myr.\nextline
Since we are using a known massive star catalogue, the reported fractions strictly apply to the targets in the ALS II catalogue, and not to the full population of OB+BH systems in the Milky Way. As the true distance distribution is more skewed towards larger distances, many more systems would be too faint to be detected and hence the detection fraction would be lower.
\begin{figure*}
    \centering
    \includegraphics[width = \textwidth]{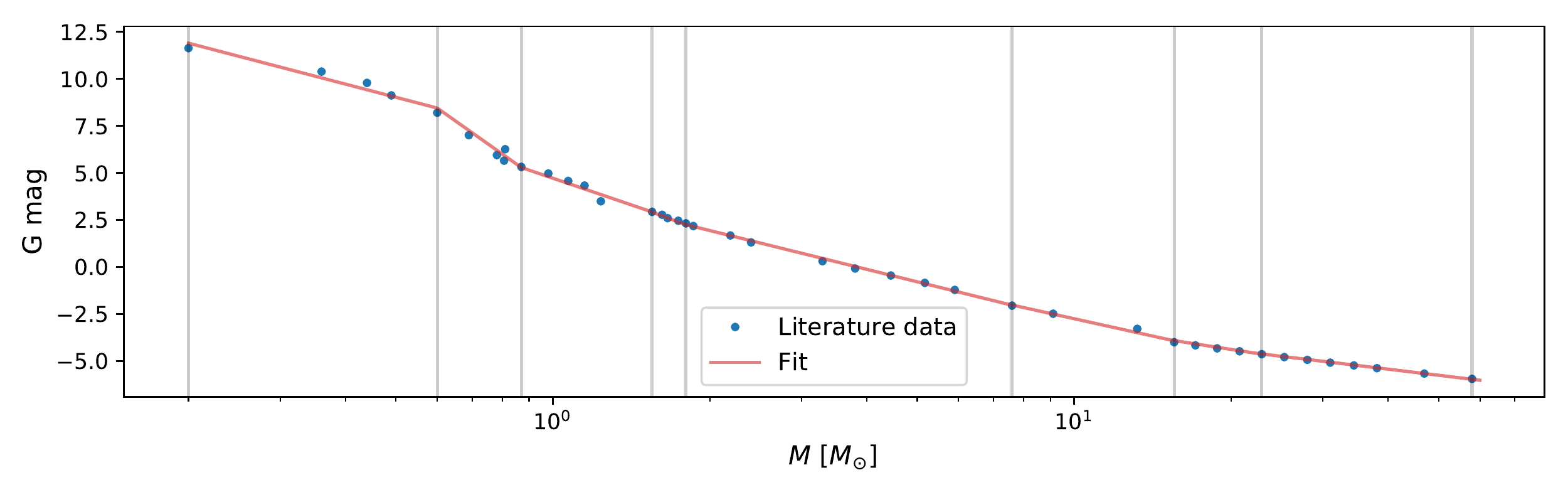}
    \caption{$G$-band magnitudes for dwarfs in the mass range 0.2-55.67\,$\Modot$ obtained from literature (see \Tabref{table_data_mass-mag}). The masses are shown on a logarithmic scale. The fit is shown with a red solid line. Grey vertical lines indicate the fitted regions.}
    \label{fig_mass-magnitude}
\end{figure*}
\subsection{Apparent $G$-band magnitudes}\label{sec_G-magnitudes}
\begin{figure*}
    \centering
    \begin{subfigure}{0.498\linewidth}
    \includegraphics[width = \textwidth]{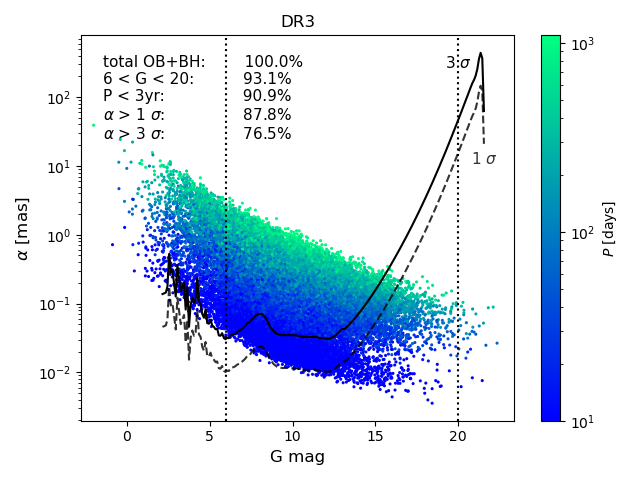}
    \end{subfigure}
    \vspace{-0.3in}
    \hspace{-0.1in}
    \begin{subfigure}{0.498\linewidth}
    \includegraphics[width = \textwidth]{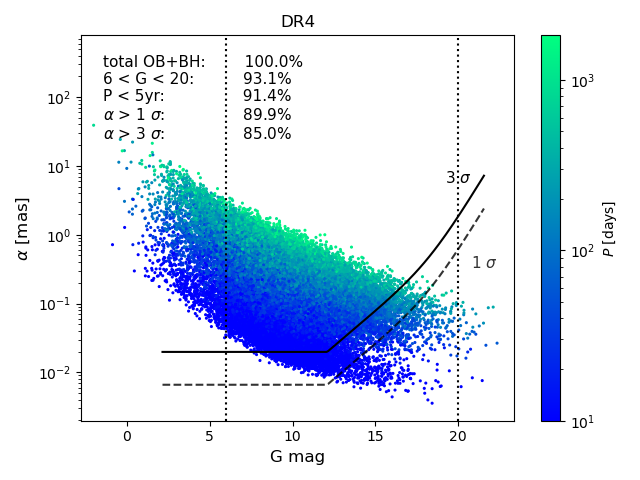}
    \end{subfigure}
    \begin{subfigure}{0.498\linewidth}
    \includegraphics[width = \textwidth]{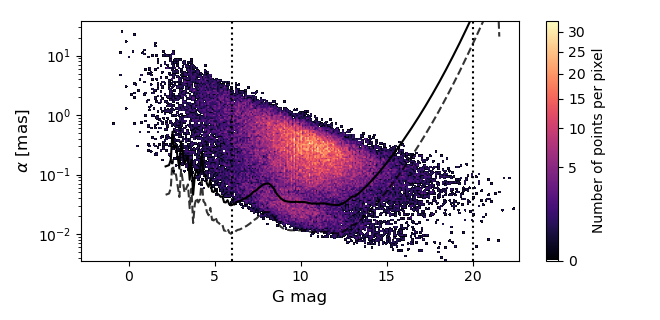}
    \end{subfigure}
    \hspace{-0.1in}
    \begin{subfigure}{0.498\linewidth}
    \includegraphics[width = \textwidth]{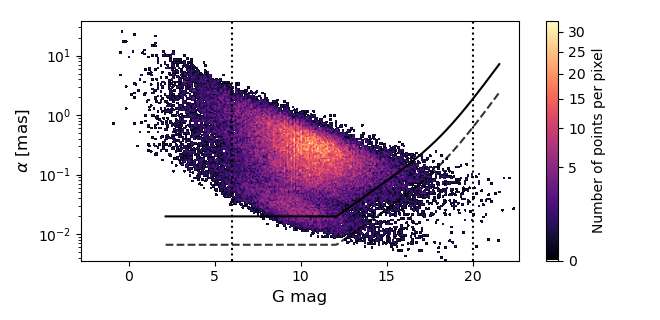}
    \end{subfigure}
    \caption{Astrometric signal and magnitudes of the simulated OB+BH systems, left for DR3 and right for DR4. Top panels are with a color-coded period, bottom panels are density plots. Systems with periods larger than the upper limit of the respective data release are not shown. The vertical dotted lines in the panels show the limiting \textit{Gaia} magnitude. From top to bottom, the meaning of the different entries in the legend is the fraction of OB+BH systems in total (per definition 100\%), with $G$-band magnitude between 6 and 20 mag, with a period below the upper limit, and that have an astrometric signal larger than 1 or 3 $\sigma$, respectively. Systems that pass a certain criterion also need to have passed the criteria mentioned above it. The two higher-density regions in the bottom plot originate in the underlying period distribution related to the mass transfer event \citep[see also the upper panel of Fig. 6 in][]{Langer_2020}. Systems undergoing Case A mass transfer are found at shorter periods (the bottom region) and systems going through Case B mass transfer end up with higher periods (upper region).}
    \label{fig_astrometric_signal_Gaia}
\end{figure*}

For reliable \Gaias astrometric solutions, we applied a conservative magnitude limit of $6<G<20$ based on EDR3 data and performance \citep{Lindegren_2020}. We determined the apparent $G$-band magnitudes $m_G$ of the LCs in the OB+BH systems as
\begin{equation}\label{eq_magnitudes}
    m_G = M_G + 5 \log d - 5 + A_G d\times 10^{-3},
\end{equation}
where $M_G$ is the absolute $G$ magnitude, $d$ is the distance to the system in parsec, and $A_G$ is the extinction in the $G$ band per kpc. The true situation of extinction is complex and depends not only on distance, but also on position in the sky. For the simplified approach, however, an average constant extinction can be used. Using table 3 from \citet{Wang_2019} and an average $V$-band extinction $A_V = 1$\,mag kpc$^{-1}$ for the Milky Way \citep[see][]{Spitzer_1978}, we obtain $A_G = 0.789$\,mag kpc$^{-1}$. \nextline 
We determined the absolute $G$-band magnitude as the difference between the bolometric magnitude and the bolometric correction factor for the $G$-band, or $M_G = M_{\text{bol}} - \text{BC}_G$. The bolometric magnitude is calculated as $M_{\text{bol}} = -2.5\log(L/L_{\odot}) + M_{\text{bol}, \odot}$, where we adopt a solar value $M_{\text{bol}, \odot}$ of 4.74.
The luminosity $L$ of an LC is determined from Stefan-Boltzmann's law. We used the bolometric correction factors available on the MIST website\footnote{\url{http://waps.cfa.harvard.edu/MIST/model_grids.html\#bolometric}} \citep[MESA Isochrones and Stellar Tracks;][]{Choi_2016, Dotter_2016} . The correct bolometric correction factor is determined by the effective temperature and the surface gravity ($g = \text{G}M/R^2$).\nextline
The obtained absolute and apparent magnitudes of the simulated OB+BH systems (see \Secref{sec_BH_distributions_title}) are shown in the right panel of \Figref{fig_distances_magnitudes}. The apparent and absolute (calculated with Eq. \ref{eq_magnitudes}) magnitudes of the massive stars in ALS II are also shown as well as the imposed magnitude limits. We did not include any correlation between mass and distance when assigning distances in \Secref{sec_distances}. The longest distances in the ALS II most likely correspond to only the most massive stars in the catalogue. Similarly, at the shortest distances lower-mass massive stars are found. Although this has not been taken into account, the closeness of the apparent magnitudes of the simulated sample and the ALS II shows that this correlation does not impact our results significantly. If any, the impact of the slight overabundance of fainter systems in the simulation is smaller detection fractions as the \Gaias precision is lower for fainter stars (see \Figref{fig_astrometric_signal_Gaia}).

\begin{table}
\centering
\caption{Fit parameters for the mass-magnitude relation of dwarfs in different mass regimes.}
\begin{tabu}{ ccc }
     \hline
     \hline
     $M_{\text{low}}-M_{\text{up}}$ [$\Modot$] & $a$ & $b$\\ 
     \hline
     22.9 - 57.95 & $-3.31 \pm 0.08$ & $-0.12 \pm 0.11$\\ 
     15.55 - 22.9 & $-4.22 \pm 0.43$ & $1.12 \pm 0.59$\\
     7.6 - 15.55 & $-6.07 \pm 0.25$ & $3.33 \pm 0.31$\\
     1.8 - 7.6 &$-6.83 \pm 0.06$ &  $4.00 \pm 0.12$\\
     1.55 - 1.8 & $-10.51 \pm 5.66$ & $4.93 \pm 1.45$\\
     0.87 - 1.55 & $-9.41 \pm 0.42$ & $4.73 \pm 0.39$\\
     0.6 - 0.87 & $-19.58 \pm 0.93$ & $4.11 \pm 0.40$\\
     0.02 - 0.6 & $-7.23 \pm 0.21$ & $6.85 \pm 0.43$\\
     \hline
     \end{tabu}
\label{table_G-mag_fit}
\end{table}

\subsection{The astrometric signal}\label{sec_astrometric_signal}
For DR3, we require a conservative upper limit on the orbital period $P\,\leq\,3$\,yr, such that at least one full period is covered by the data. In an extension to DR4, the maximum allowed period is $5$\,yr. A lower limit is determined by the astrometric signal of the binary system, which is equal to the observed angular semi-major axis of the photocentre motion. In order for \textit{Gaia} to detect a source as a binary, we require that the observed astrometric signal is significantly larger than the \textit{Gaia} astrometric precision, $\alpha \geq (3) \sigma_G$ as a (conservative) criterium.\nextline
The \textit{Gaia} astrometric precision is magnitude dependent. For DR3, we can use the precision measured with EDR3 data. We used for the EDR3 \textit{Gaia} precision the derived along-scan variance data provided to us by A. Everall \citep[priv. comm., based on a similar method described in][]{Everall_2021}. For the precision at the end of the nominal 5-year mission, we used the estimated end-of-mission precision as described in Sec. 8.1 (eqs. 4,5,6) of \citet{Gaia_colab_2016_mission}.\\
\nextline
We determine an expression for the astrometric signal produced by the binary system, which is related to the semi-major axis of the projected orbit $\upalpha$ and the distance $d$ to the system through $\alpha = \upalpha/d$. In binary systems with a non-luminous component, the observed photocentre motion is equal to the motion of the LC around the common centre of mass. Its semi-major axis in the true orbit is given by $a_{\text{LC}} = a M_{\text{BH}}/M_{\text{tot}}$, where $a=a\LC + a\BH$ is obtained from Kepler's third law $4\pi^2 a^3 = \text{G}M_{\text{tot}}P^2$. The projected orbit of the LC is then determined by the orbital parameters of orientation, $i$ (inclination), $\Omega$ (longitude of the ascending node), and $\omega$ (argument of periastron), through the \textit{Thiele-Innes} constants
\begin{equation}\label{eq_Thiele-Innes}
    \begin{split}
        A &= a_{\text{LC}} \left[ \cos\Omega\cos\omega - \sin\Omega\sin\omega\cos i   \right],\\
        B &= a_{\text{LC}} \left[ \sin\Omega\cos\omega + \cos\Omega\sin\omega\cos i   \right], \\
        F &= a_{\text{LC}} \left[ -\cos\Omega\sin\omega - \sin\Omega\cos\omega\cos i   \right],\\
        G &= a_{\text{LC}} \left[ -\sin\Omega\sin\omega + \cos\Omega\cos\omega\cos i   \right].
    \end{split}
\end{equation}
For the systems, $\Omega$ and $\omega$ were uniformly randomly distributed between $0$ and $2\pi$, and $\cos i$ between $-1$ and $1$. 
The projected Cartesian coordinates $x_{\text{proj}}$ and $y_{\text{proj}}$ are given by
\begin{equation}\label{eq_projected_coordinates}
    \begin{split}
        x_{\text{proj}} &= AX + FY,\\
        y_{\text{proj}} &= BX + GY,
    \end{split}
\end{equation}
where $X = \cos E - e$ and $Y = \sqrt{1-e^2} \sin E$, with $E$ being the eccentric anomaly. \nextline 
For eccentric systems, the semi-major axis $a$ of the orbit, hereafter called the true semi-major axis, and the semi-major axis of the projected orbit $\upalpha$ generally do not have the same length. An example is shown in \Figref{fig_projected_semimajor-axis}. For this eccentric system (see Figure caption for more details), the projection of the true semi-major axis has a length of 0.75\,au, while the semi-major axis of the projected orbit is $\upalpha_{\text{LC}} = 1.00$\,au. 
However, for non-eccentric systems ($e=0$), we have $a_{\text{LC}} = \upalpha_{\text{LC}} = a_{\text{LC,proj}}$, no matter the orientation of the system. (A derivation of the equality together with an expression for $\upalpha$ for eccentric orbits is given in Appendix \ref{app_semi-major_axis_proj}).\nextline In our simulations, we use the observational constraint of $\alpha>3\sigma_G$, ignoring the orbital inclination. However, the observed astrometric {\it signal} of the orbital motion, which determines the detectability of the orbit, is the actual motion projected on the plane of the sky, and therefore depends on the orbital inclination. For inclinations larger than $0^{\degree}$, the projection of one of the two dimensions of the orbital motion is suppressed by $\cos i$, and therefore the effective astrometric signal, which is used as the observational cut, is, on average, suppressed by $\sqrt{(1+\cos^2 i)/2}$, a correction factor we applied in the simulations.

\begin{figure}
    \centering
    \includegraphics[width = 0.495\textwidth]{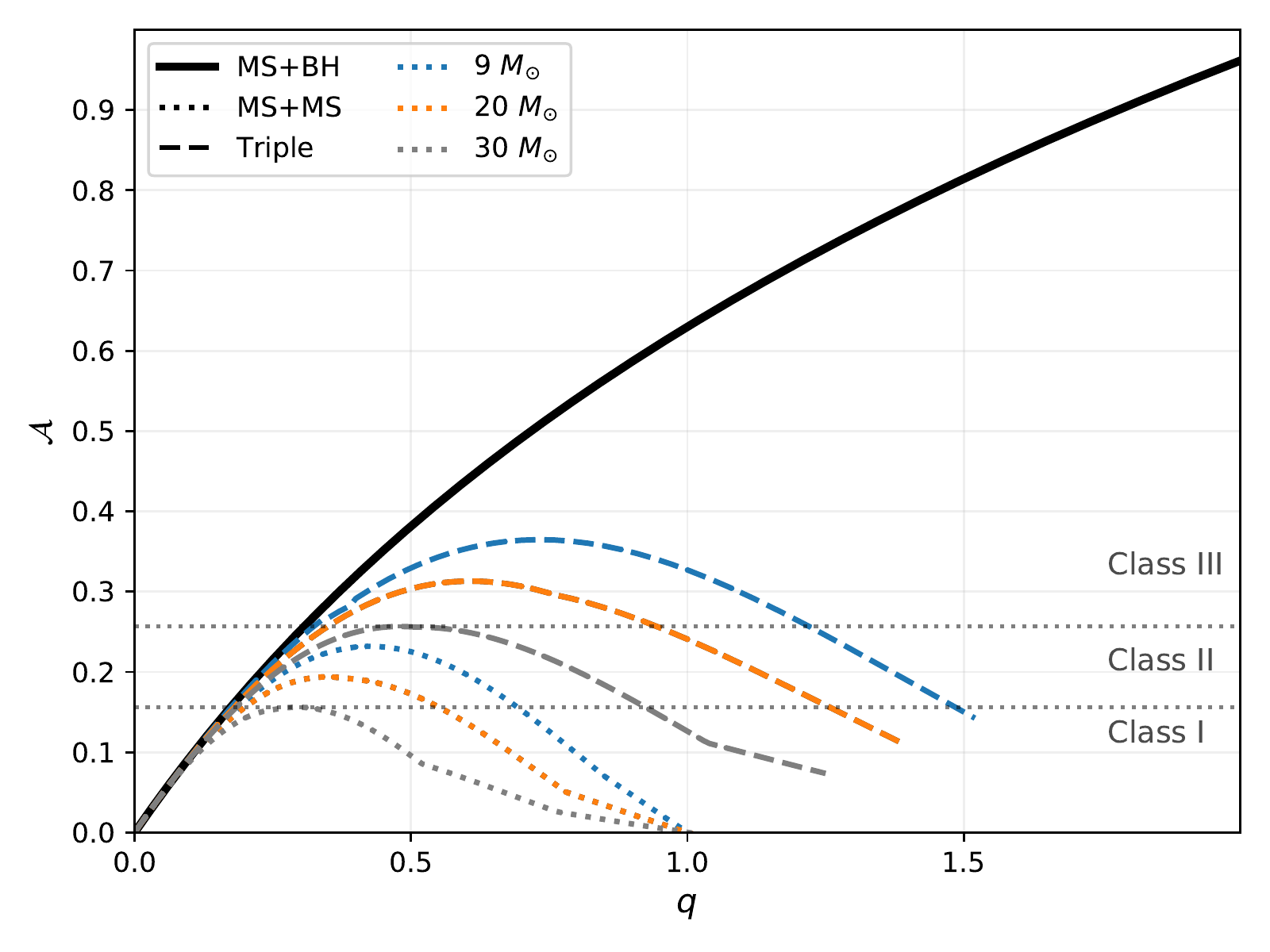}
    \caption{Different theoretical AMRFs for three different primary masses of $9\Modot$ (blue), $20\Modot$ (orange), and $30\Modot$ (gray). The black solid line shows the theoretical AMRF for MS+BH systems, the dotted curves are for main-sequence binaries, and dashed curves are for triple main-sequence systems. Different classes are also indicated for the $30\Modot$ primary by the horizontal dotted lines at the maxima of the triple-AMRF and the main-sequence-AMRF (Class III: identifiable OB+BH systems, Class II: 
    triple main-sequence or OB+BH systems, Class I: OB+OB binaries, triple main-sequence systems, or OB+BH binaries).}
    \label{fig_shahaf_empty}
\end{figure}

\section{Identifying the OB+BH systems}\label{sec_shahaf_title}
The next step is to separate the needles -- the OB+BH systems -- from the haystack -- normal binaries with two main-sequence components.
For this, we used the method described by \citet{Shahaf_2019}. Their method astrometrically identifies binaries with a compact object, based on the observed astrometric signal of the binaries. 
\citet{Shahaf_2019} define a dimensionless parameter called the Astrometric Mass-Ratio Function (AMRF or $\mathcal{A}$), defined as 
\begin{equation}\label{eq_AMRF_general_binary}
    \mathcal{A} = \frac{\alpha}{\varpi} \left( \frac{M_1}{M_{\odot}} \right)^{-1/3} \left( \frac{P}{\text{yr}} \right)^{-2/3} = \frac{q}{(1+q)^{2/3}} \left( 1- \frac{S(1+q)}{q(1+S)}\right),
\end{equation}
where $\varpi$ is the parallax of the system, and $q = M_2/M_1$ and $S = I_2/I_1$ are the current mass ratio and intensity ratio, respectively, of the secondary (least luminous) to the primary (most luminous) star. The intensity of a star in the $G$ band can be defined as $I^G \propto 10^{-0.4M_G}$ and the intensity ratio between two components hence becomes $S^G(M_1, M_2) = 10^{-0.4\Delta M_G}$. This expression does not take into account the age of the stars. From the mass-magnitude relation (established in \Secref{sec_mass-magnitude_relation}) we can determine the intensity ratio for a given primary mass at different mass ratios and determine the theoretical AMRF (\Secref{sec_theoretical_AMRF}).\nextline
In \Eqref{eq_AMRF_general_binary}, we will call the expression with $\alpha$ the observational AMRF or just $\mathcal{A}$ and the right-most expression the theoretical AMRF. As \citet{Shahaf_2019} showed, the AMRF is a useful parameter for separating BH systems from non-degenerate systems, as we also show in \Secref{sec_Shahaf_observed}. \nextline

\subsection{The mass-magnitude relation in the $G$ band.} \label{sec_mass-magnitude_relation}
\citet{Shahaf_2019} have used their method on a \textit{Hipparcos} sample of low-mass dwarf stars, establishing a piecewise linear mass-magnitude relation of the form $M_G = a \log(M/\Modot) + b$. Similarly, we derived such a logarithmic mass-magnitude relation for the \Gaias $G$-band magnitude. \Tabref{table_data_mass-mag} lists the collected temperatures and radii for dwarfs with masses in between 0.20-57.95\,$\Modot$, covering spectral types O3\,V\,-\,M9\,V. Using the same method as described in \Secref{sec_G-magnitudes}, we obtained the $G$-band magnitudes, which are shown in \Figref{fig_mass-magnitude}.\nextline
To establish the piecewise linear mass-magnitude relation, we divided the obtained magnitudes in different regions based on their mass. \Tabref{table_G-mag_fit} shows the fitted regions with their respective fit parameters and the fit is shown in \Figref{fig_mass-magnitude}.

\subsection{The theoretical AMRF curves for different masses}\label{sec_theoretical_AMRF}
\citet{Shahaf_2019} distinguish between different types of systems, based on the nature of the secondary star. The primary star is defined as the most massive LC (for OB+BH systems, the LC is always the primary star). Three different types of systems are defined: main-sequence binaries, triple main-sequence systems, where the secondary star is an unresolved main-sequence binary, and MS+BH binaries. For each of these cases, the same value of $q$ will lead towards different intensity ratios $S$ and hence \Eqref{eq_AMRF_general_binary} will appear different. We highlight that $q$ is always defined as the mass of the companion (either a BH, another main-sequence star, or a binary system) divided by the mass of the primary main-sequence star.

\begin{figure}
    \centering
    \begin{subfigure}{0.495\linewidth}
    \includegraphics[width = \textwidth]{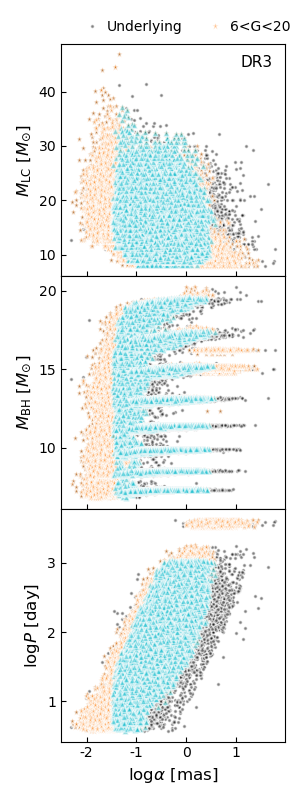}
    \end{subfigure}
    \hspace{-0.2cm}
    \begin{subfigure}{0.495\linewidth}
    \includegraphics[width = \textwidth]{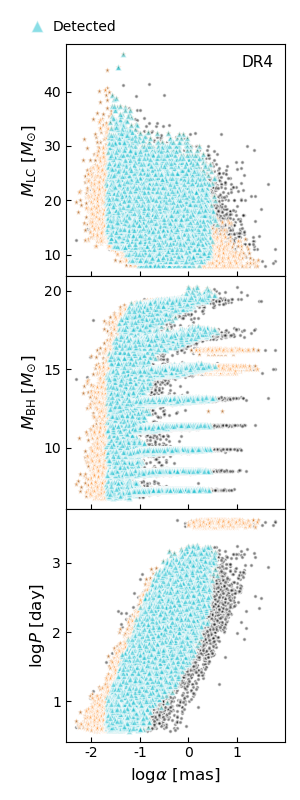}
    \end{subfigure}
    \caption{Masses of the components and the periods as a function of the astrometric signal. The underlying distributions \citep[those of][]{Langer_2020} are shown with black dots and overplotted with those systems having $6<G<20$ (orange stars). On top is the subset of systems that are detected in the conservative case (i.e. $\alpha > 3\sigma$ and a period lower than the imposed upper limit, cyan triangles). Left for DR3, right for DR4. The discrete BH masses come from the discrete main-sequence primary masses in the simulations of \citet{Langer_2020}.
    }
    \label{fig_dist_Gmag_vs_...}
\end{figure}

Figure \ref{fig_shahaf_empty} shows theoretical AMRFs obtained with the right-most term in \Eqref{eq_AMRF_general_binary}, using the mass-magnitude relation derived in \Secref{sec_mass-magnitude_relation}. We show curves for three different primary masses. The black solid line shows the theoretical AMRF for OB+BH systems. Dotted curves are for main-sequence binaries, and the dashed curves are for triple systems for which the binary companion has two equal-mass components. The triple curves are cut-off when the binary component would become brighter than the primary star.\nextline
The dotted horizontal lines show the maximum of the theoretical AMRF for main-sequence binaries and triple main-sequence systems with a $30\Modot$ primary. Based on these maxima, \citet{Shahaf_2019} distinguish between three classes (also indicated on the figure for a $30\Modot$ primary). Systems with an observational $\mathcal{A}$ above the maximum of the triple-AMRF (corresponding to 0.365, 0.313, and 0.257 for the 9, 20, and 30\,$\Modot$ primary, respectively) are classified as Class III. 
These system are identifiable OB+BH systems. Systems with $\mathcal{A}$ above the maximum of the main-sequence-AMRF (corresponding to 0.232, 0.194, and 0.156 for the 9, 20, and 30\,$\Modot$ primary, respectively) and below the maximum of the triple-AMRF are classified as Class II. Most of these systems are triples, however some might be OB+BH systems. Finally, Class I are those systems with an $\mathcal{A}$ below the maximum of the main-sequence-AMRF and hence the nature of these objects can be either a main-sequence binary (most probable), a triple main-sequence system, or an OB+BH system.\nextline
Although OB+NS systems might also be present in the observed sample (i.e. in \Gaia), they will most likely not be identified as Class III. For $q$ values in the range of NS companions with massive OB-type LCs ($q \sim 0.05-0.3$), all curves lie close to each other and hence we would not be able to distinguish between the three proposed cases.
\begin{figure*}
    \centering
    \begin{subfigure}{0.497\linewidth}
    \includegraphics[width = \textwidth]{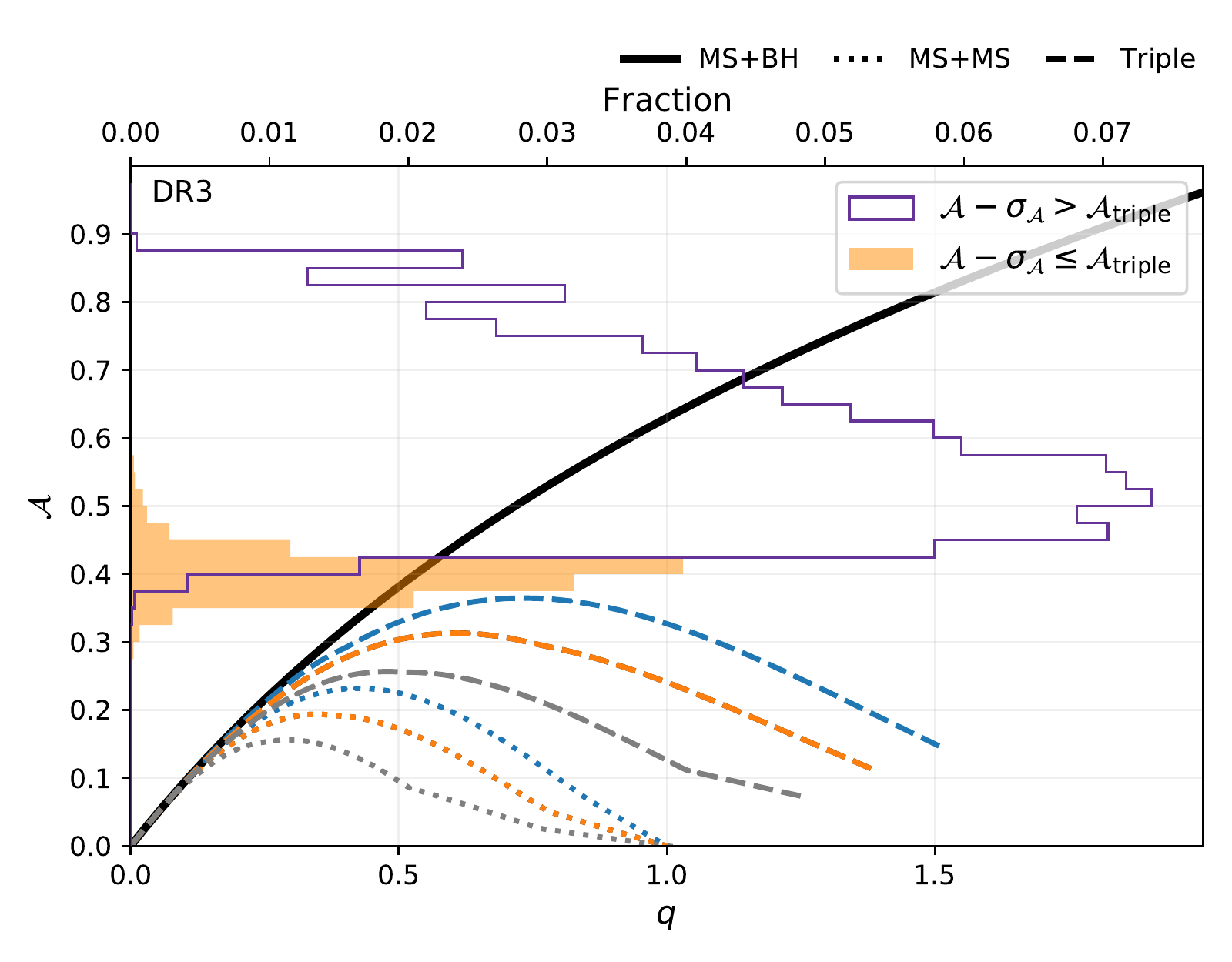}
    \end{subfigure}
    \begin{subfigure}{0.497\linewidth}
    \includegraphics[width = \textwidth]{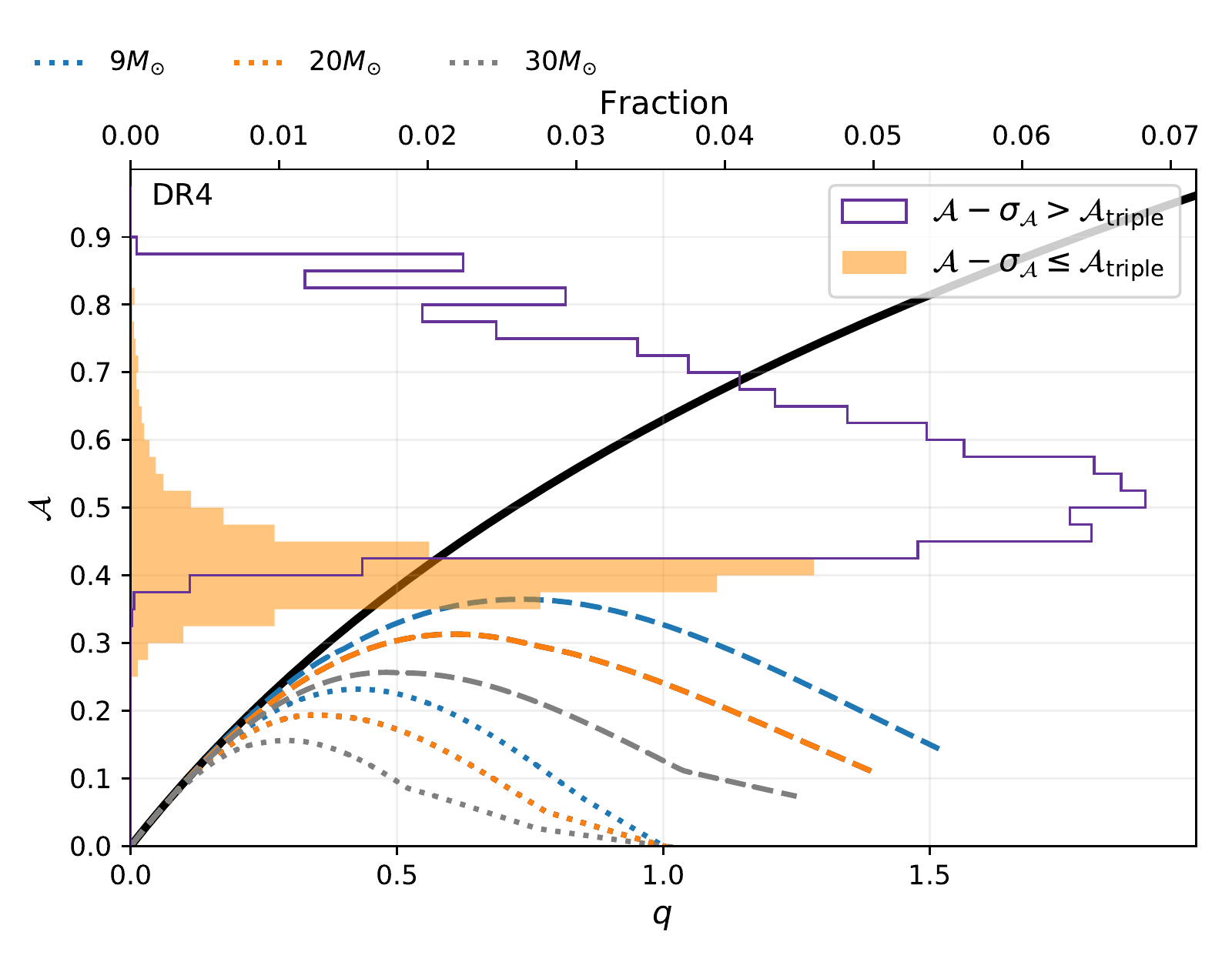}
    \end{subfigure}
    \caption{The distribution of $\mathcal{A}$ for the systems classified as a binary by \Gaia. The left panel is for DR3 and the right panel for DR4. The curves are the same as in \Figref{fig_shahaf_empty} (legend on top). Systems with $\mathcal{A}-\sigma_{\mathcal{A}} > \mathcal{A}_{\text{triple}}$ (Class III) are represented by the purple open histogram, systems with $\mathcal{A}-\sigma_{\mathcal{A}} \leq \mathcal{A}_{\text{triple}}$ (non-Class III) are shown with an orange filled histogram. Bins are 0.025 in width. The open histograms account for 88.7\% and 82.3\% of the total number of detected systems in the left and right panel, respectively.}
    \label{fig_shahaf_observed}
\end{figure*}
\begin{figure*}
    \centering
    \begin{subfigure}{0.495\linewidth}
    \includegraphics[width = \textwidth]{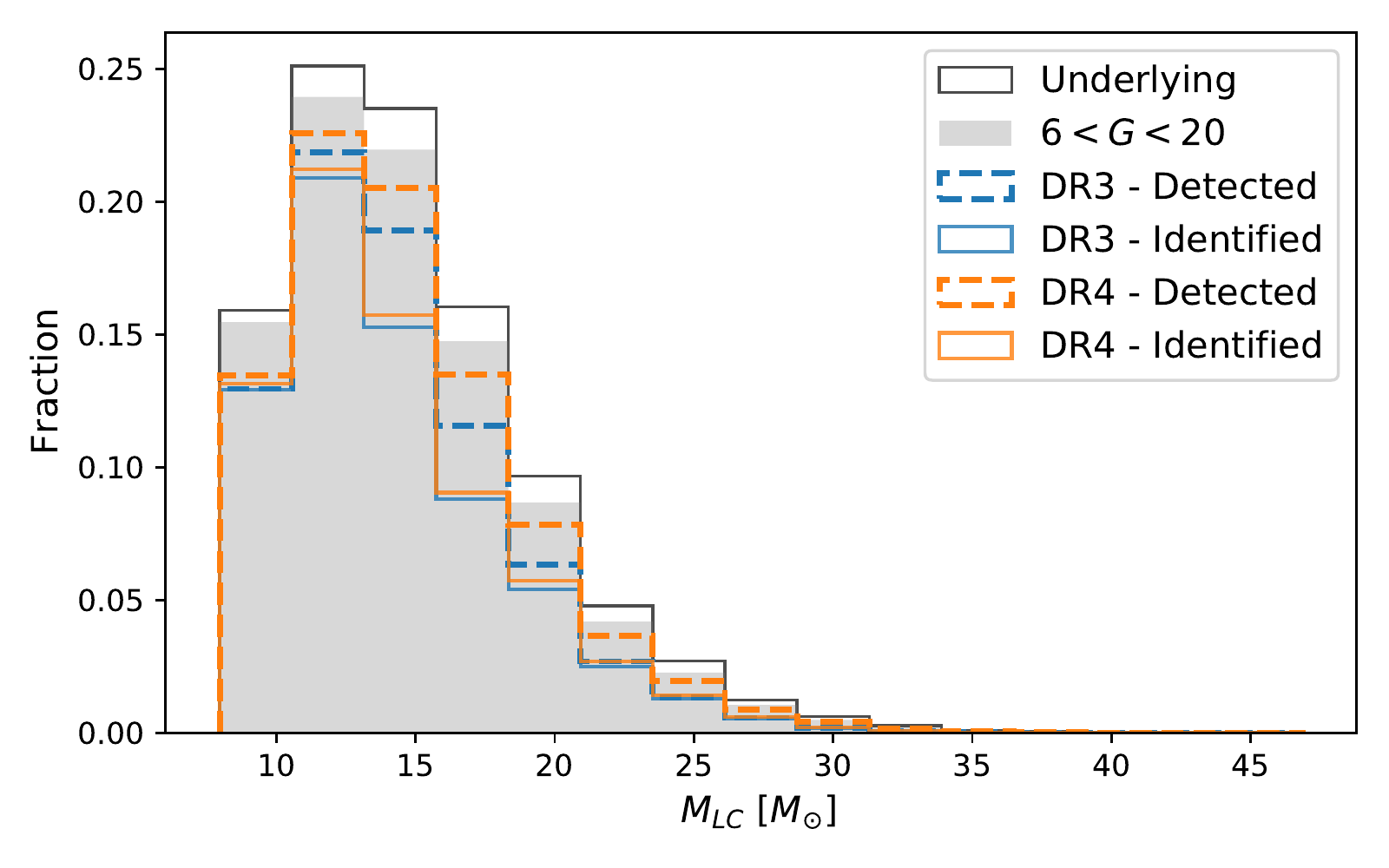}
    \end{subfigure}
    \begin{subfigure}{0.495\linewidth}
    \includegraphics[width = \textwidth]{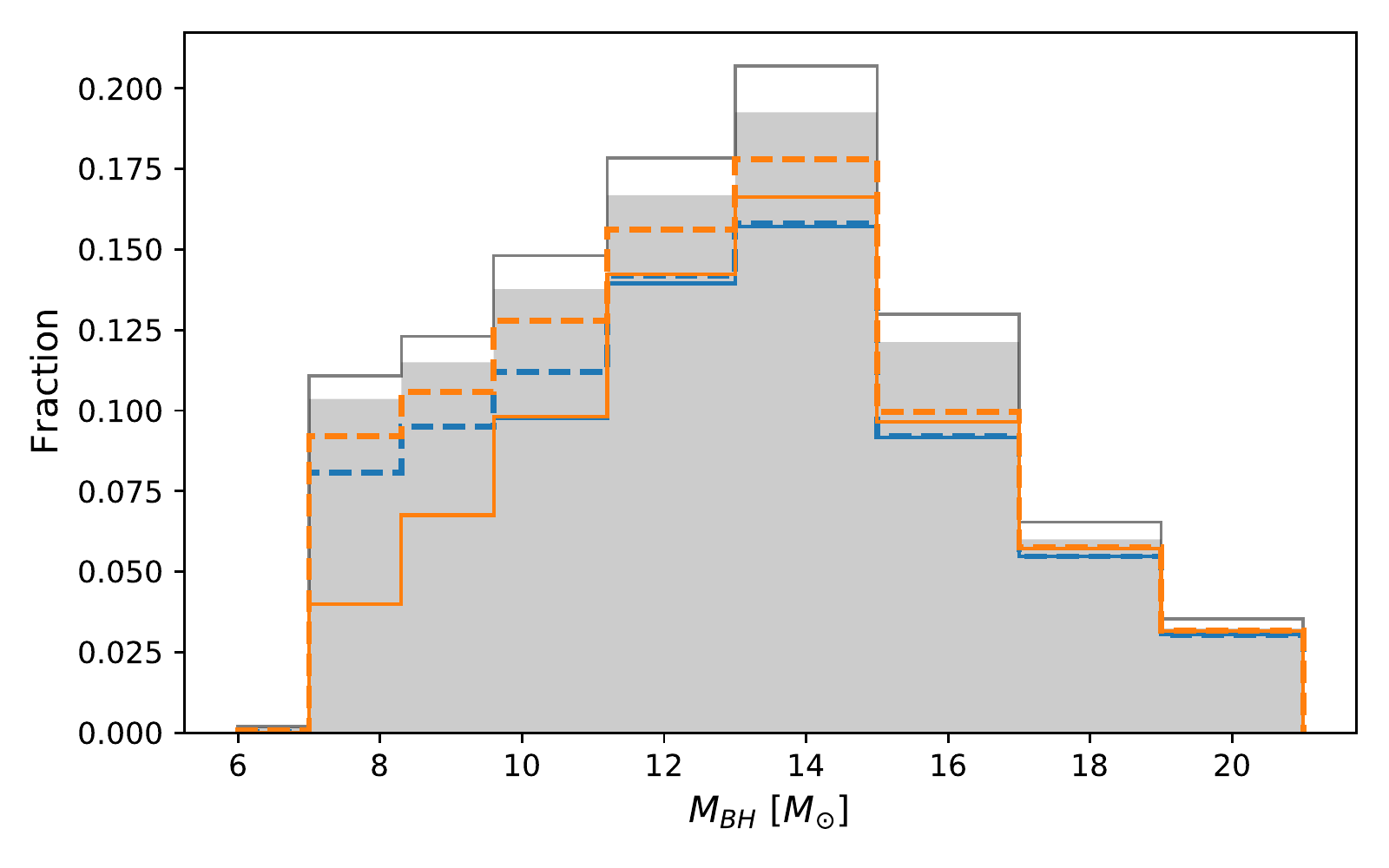}
    \end{subfigure}
    \begin{subfigure}{0.495\linewidth}
    \includegraphics[width = \textwidth]{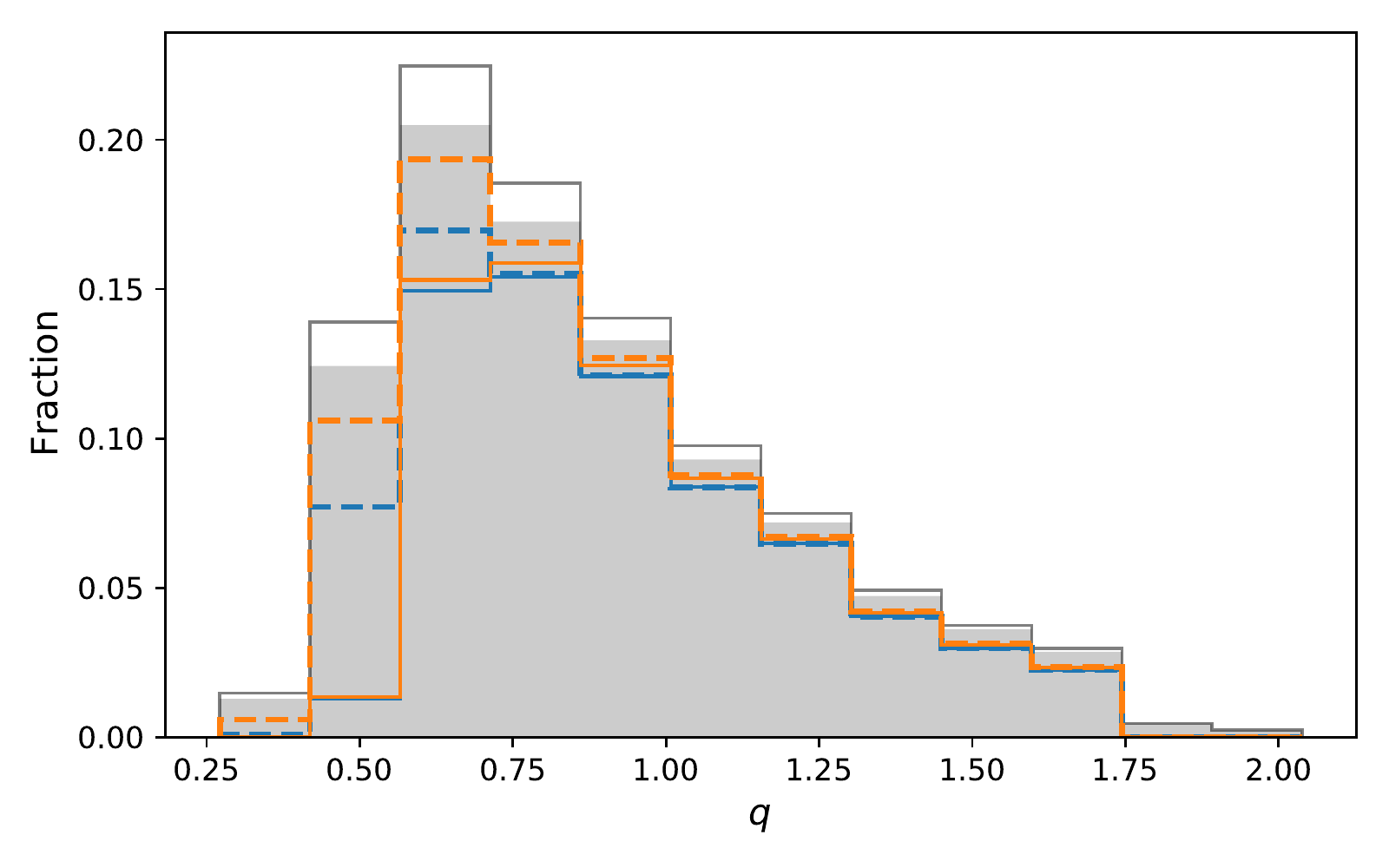}
    \end{subfigure}
    \begin{subfigure}{0.495\linewidth}
    \includegraphics[width = \textwidth]{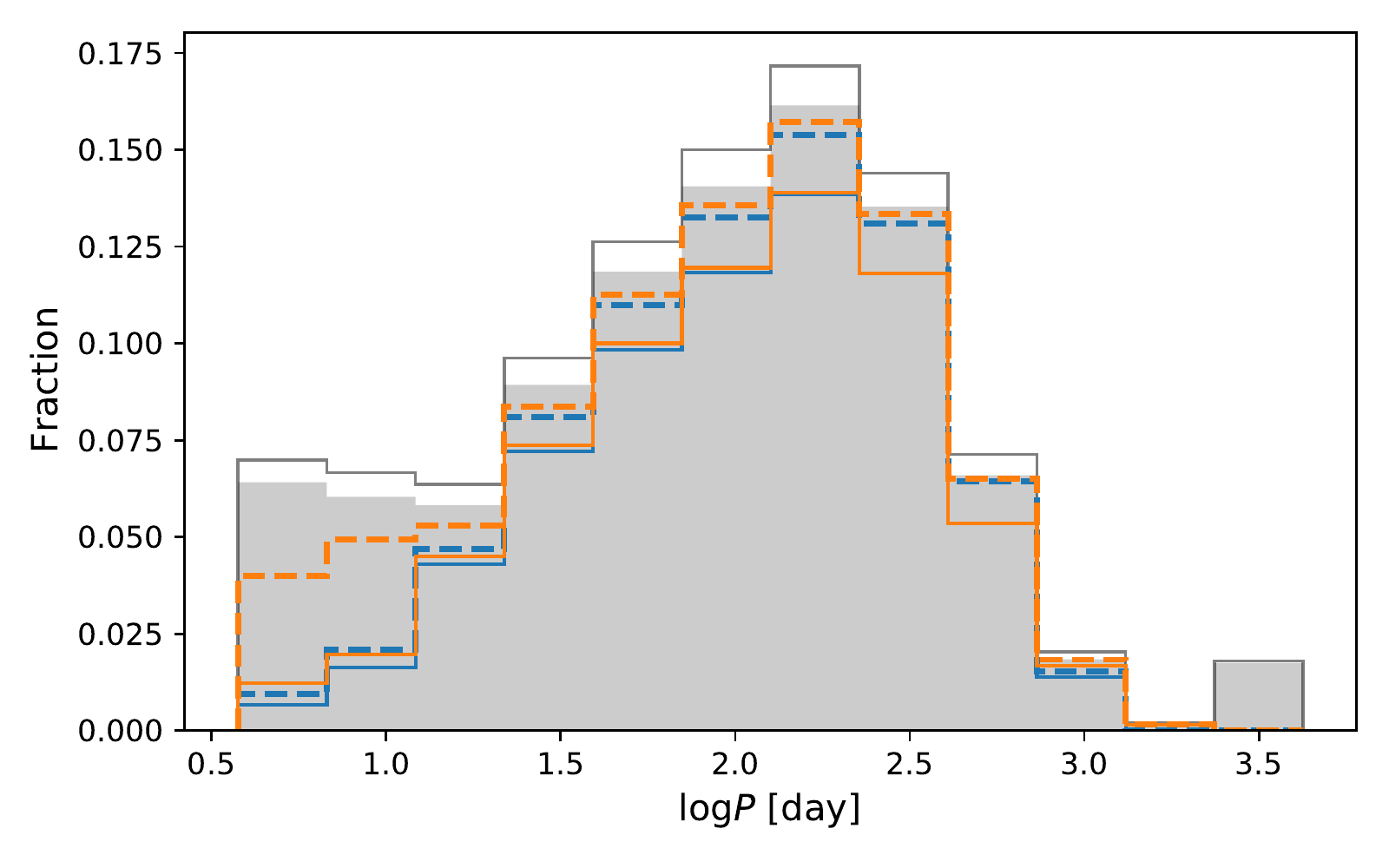}
    \end{subfigure}
    \caption{Distribution of $M_{\text{LC}}$ (top left), $M\BH$ (top right), $q$ (bottom left), and $P$ (bottom right) in DR3 and DR4. In the legend, `Underlying' is the distribution of all simulated OB+BH systems, `Detected' are those systems detected as binaries by \Gaia, and `Identified' systems are Class III systems (i.e., confirmed OB+BH systems). The latter are the observational distributions that can be expected to be constructed using the OB+BH systems identified in DR3.}
    \label{fig_observed_distributions}
\end{figure*}

\section{The detected OB+BH distribution}\label{sec_BH_distributions_title}
To predict the theoretical distribution of the masses of the components and orbital periods of the OB+BH systems that will be identified, we performed a statistical analysis by drawing 50\,000 OB+BH systems from the simulations, using a weight factor depending on the lifetime of the systems and the observed distributions of the initial period and masses, while assuming a constant star formation rate \citep[see][]{Langer_2020}. This way, we can estimate the fraction of OB+BH systems that can be detected and identified. Therefore, we will only report results as fractions in this Section. Later, we can apply these predicted detection fractions to the expected number of identifiable OB+BH systems in the ALS II (see \Secref{sec_estimated_numbers}). \nextline
The reported fractions in this Section are based on the assumption that the masses of the LCs are known. The effect of uncertainties on the masses is investigated in \Secref{sec_uncertainties}.

\subsection{The fraction of OB+BH systems detected by \Gaia}\label{sec_Gaia_OB+BHs}
We first investigated how many OB+BH systems \Gaias would be able to detect as binaries, using the observational constraints presented in \Secref{sec_Gaia_cuts_title}.
The results are presented in \Figref{fig_astrometric_signal_Gaia}. \nextline
Accounting for the \Gaias magnitude range, 93.1\% of our simulated OB+BH binaries would be included as sources in the \Gaias catalogue. Using the conservative observational limit ($\alpha > 3\sigma$), 82.2\% of those \Gaias sources would be flagged as binaries in DR3, amounting to 76.5\% of the entire sample. At the end of the 5-yr mission, 85.0\% of the entire sample will be detected as binaries. Below, we adopt these conservative detection estimates.
\nextline
The lower limit imposed on the astrometric precision mostly filters out (the closest) systems with the shortest periods. Although in line with what we expect, it is worth noting that the observed population of OB+BH systems will be slightly biased towards systems with larger orbital periods. However, systems with the largest orbital periods are also excluded from detection due to the imposed upper limit on the period. Plots for the observed parameter space are shown in \Figref{fig_dist_Gmag_vs_...} for the masses of the components and the orbital periods with the astrometric signal and corner plots of the distributions for different observable parameters are shown in \Figrefs{fig_corner_3yr} and \ref{fig_corner_5yr} for DR3 and DR4, respectively, for the full sample, the \Gaias sources, and the sources detected as binaries. The comb-like structures in the plots of the BH mass as a function of, for example, the period or astrometric signal originate in the discreteness in the initial parameters of the binary model simulation. However, Case A mass transfer -- mass transfer while both components are still on the main sequence -- in shorter period systems results in the fading of the discreteness in the distribution of the parameters of the post-mass transfer systems and hence the disappearance of the comb-like structures.

\subsection{Finding the needles in the haystack: identifying the BH+OB binaries}\label{sec_Shahaf_observed}
From the fraction of OB+BH systems that will be detected as binaries by \Gaia, we investigate how many of those we can extract from the vast amount of binary sources in the \Gaias binary catalogue (see \Secref{sec_shahaf_title} for the method). To only consider significant identifications, we determine if the measured value of $\mathcal{A}-\sigma_{\mathcal{A}}$ of a binary system is larger than the maximum $\mathcal{A}_{\text{triple}}$ reached by the triple AMRF corresponding to the mass of the LC. If this is the case, we count that system classified as an OB+BH system. We used the orbital period and the mass of the LC from the simulated systems, and their calculated astrometric signal to calculate $\mathcal{A}$ (Eq. \ref{eq_AMRF_general_binary}) for each system. The uncertainty $\sigma_{\mathcal{A}}$ on $\mathcal{A}$ was calculated using the \Gaias precision as an uncertainty on the parallax and astrometric signal, a 10\% uncertainty on the period, and a 50\% uncertainty on the mass.\nextline
Figure \ref{fig_shahaf_observed} shows how many of the systems classified as a binary by \Gaias have $\mathcal{A}-\sigma_{\mathcal{A}} > \mathcal{A}_{\text{triple}}$ and are hence classified as Class III systems. For DR3, 88.7\% of the OB+BH systems classified as binaries are classified as Class III systems in this diagram, hereafter called the AMRF completeness. Combining this AMRF completeness with the detectable fraction, we find that we are able to identify 67.8\% of all sampled OB+BH systems. For DR4, we have an AMRF completeness of 82.3\%, leading to the identification of 69.9\% of all sampled OB+BH systems.

\subsection{Distributions of the orbital period and masses}\label{sec_observed_distributions}
We investigate the expected distributions of the masses of the components in the identified OB+BH systems and their periods. We can determine three things: 1) the expected distributions of the systems which are observed by \Gaia, 2) those of the systems which are detected as binaries by \Gaia, and, more interestingly, 3) the expected distributions of the identifiable OB+BH systems. Figure \ref{fig_observed_distributions} shows the distributions of the mass of the LC and BH, the mass ratio, and the period in DR3 and DR4.
\nextline 
The systems with the shortest and longest periods are not detected by \Gaias as binaries in DR3. The same is seen for systems with the lowest and highest mass ratios. In DR4, a significant fraction of the short-period and low-mass-ratio systems will be detected by \Gaia. The longest-period systems are not detected in any of the releases due to our upper limit imposed on the period. These are also the systems with the largest mass ratios. Then, we see that in the identification of the OB+BH systems, we will identify relatively more of the massive ($\gtrsim 13\Modot$) BHs rather than the lower mass ones, resulting in relatively more systems with mass ratios above 0.8. Also the shortest period systems will not be identified. These biases will not disappear in DR4.

\section{Discussion}\label{sec_discussion}
\subsection{The number of identified OB+BH binaries}\label{sec_estimated_numbers}
We used the massive stars in ALS II \citep[][]{Gonzalez_2021} and the prediction that 3\% of massive OB stars in binaries have a BH companion \citep{Langer_2020}. The number of massive stars in the ALS II within our imposed magnitude range is 13\,288. Applying an initial binary fraction of 70\% in the range of $\log P/\text{d} < 3.5$ \citep[][]{Sana_2012}, there should be $\sim$280 OB+BH binaries in this sample. The conservative \Gaias detection fraction tells us that $\sim$215 of these OB+BH binaries can be detected as binaries in DR3 and $\sim$2{40} in DR4. Thanks to the high AMRF completeness, we should identify $\sim$190 of those in DR3 and $\sim$195 in DR4. The identification of this many OB+BH systems would lead to an increase in the number of known OB+BH binaries by a factor of more than 20.\nextline
The sample of known massive stars will be expanded in ALS\,III, which will use EDR3 data and many additional sources \citep[][]{Gonzalez_2021}. The estimated number of sources in ALS\,III is 17\,000. Assuming the distance distribution will be similar to that of ALS II, this would result in the identification of $\sim$240 and $\sim$245 OB+BH systems in DR3 and DR4, respectively. However, if the distance distribution differs significantly from that of ALS II, these numbers might change. \nextline
In terms of numbers, our predicted detected OB+BH systems are lower than the number of LC+BH binaries predicted in the previous works of \citet{Breivik_2017,Breivik_2019,Mashian_Loeb_2017,Yamaguchi_2018} and \citet{Wiktorowicz_2019}. The major reason for this is that we only focus on massive dwarf luminous companions instead of the full stellar population. Moreover, here we estimated numbers from an observational catalogue, which is not necessarily equal to the number of OB+BH binaries that are observed by \Gaias as it is of course not excluded that \Gaias will detect OB+BH binaries not in ALS\,II/III. However, in order to identify them as such, the spectral type or mass of the luminous companion should first be established. Furthermore, in other works it is not assumed that systems undergoing a common-envelope phase merge. Although \citet{Kruckow_2016}, \citet{Klencki_2021}, and \citet{Marchant_2021} have shown that systems undergoing a common-envelope phase are unlikely to survive, not assuming a merger could lead to more detectable systems at short periods and hence an increase in the number of identifications.\nextline
Finally, the observational constraints presented in \Secref{sec_Gaia_cuts_title} are conservative. For example, it is not excluded that \Gaias will be able to obtain orbital parameters for binary systems with $P\,>\,3$\,yr, as it is possible to derive astrometric orbital solutions without covering the full period, or $G<6$\,mag. Astrometric \Gaias solutions for these systems would only increase the number of detected and identifiable OB+BH systems. For example, considering that the brightest sources in ALS\,II would also have reliable astrometric solutions increases the number of identified OB+BH systems by about 10 in both DR3 and DR4.

\subsection{Different BH formation mechanisms}\label{sec_other_scenarios}
One of the uncertainties about the formation of BHs is related to whether or not their progenitors produce a supernova at core-collapse and whether or not a kick is introduced during BH formation. Although we only focus on the impact of the uncertainties in the BH-formation mechanism, other uncertainties are also present, such as a reduced BH formation probability in stripped stars \citep[e.g.][]{Ertl_2020, Laplace_2021,Schneider_2021}. \nextline
All previous results assumed no natal kick and that the BH progenitor undergoes a direct collapse. Here, we investigate how different BH-formation scenarios affect the identified OB+BH population.\nextline
\citet{Fryer_2012} present two explosion types: a fast and a delayed one. Fast explosions occur in the first 250 ms after the collapse has halted, due to a dramatic increase in pressure from nuclear forces and neutron degeneracy, and because the core reaches nuclear densities. A delayed explosion can happen much later. In both cases, mass can be lost in the explosion. The CO-core mass determines how much mass falls back onto the proto-compact object formed in the collapse of the progenitor, largely determining the final remnant mass. Because the simulations of \citet{Langer_2020} stop at carbon depletion and the CO-core mass can still grow further in the remaining stages, we used the relations in table 4 of \citet{Woosley_2019} to obtain the CO-core masses as a function of the He-core mass for the stars. The remnant mass was obtained using the expressions in sections 4.2 and 4.3 of \citet{Fryer_2012}.\nextline
Two additional kick scenarios were explored. The first one assumes that the formation of BHs results in kicks similar to those received by NSs. Our 3D kick velocities are Maxwellian distributed, with a 1D Gaussian distribution in the direction of each of the cartesian coordinates of the velocity vector with $\sigma_{\text{1D}} = 265$\,\kms \citep{Hobbs_2005}. The second kick scenario linearly scales the Hobbs kick velocity with the amount of fallback mass, where a larger amount of fallback mass equals a smaller kick \citep{Fryer_2012}. The extreme cases of no fallback and a complete fallback yield a Hobbs kick and no kick, respectively.\\
\nextline
We investigated 9 different BH formation mechanisms:
\begin{itemize}
    \item direct or full collapse without a kick (FC-NK), which is equal to a full collapse with a fallback-scaled NS-kick (FC-FK),
    \item full collapse with a NS-kick (FC-HK),
    \item a rapid explosion without a kick (R-NK),
    \item a rapid explosion with a fallback-scaled NS-kick (R-FK),
    \item a rapid explosion with a NS-kick (R-HK),
    \item a delayed explosion without a kick (D-NK),
    \item a delayed explosion with a fallback-scaled NS-kick (D-FK),
    \item a delayed explosion with a NS-kick (D-HK).
\end{itemize}

Both a kick and rapid mass loss through an explosion will lead to a change in orbital parameters of the systems. Hence, the eccentricity and the semi-major axis in the post-SN system will differ from those in the pre-SN system. We determined the post-SN eccentricity $e_{\text{post}}$ and semi-major axis $a_{\text{post}}$ using the equations in Chapter 10 of \citet{Pols_lectures}. 
When $e_{\text{post}} \geq 1$, the system is disrupted and hence the number of possible OB+BH detections is  decreased. In \Tabref{table_different_kicks}, the fractions of surviving OB+BH systems is listed. We have taken the same sample of pre-SN systems that was used in \Secref{sec_BH_distributions_title}. Systems which now become OB+NS systems are also excluded from the surviving systems, which is the main reason for losing more than 10\% of the simulated OB+BH systems for the R\,-\,NK scenario. Furthermore, \Tabref{table_different_kicks} also list the fractions of surviving systems that can be identified in DR3, which change due to changes in some of the parameter distributions explained below.
\begin{figure*}
    \centering
    \includegraphics[width = \textwidth]{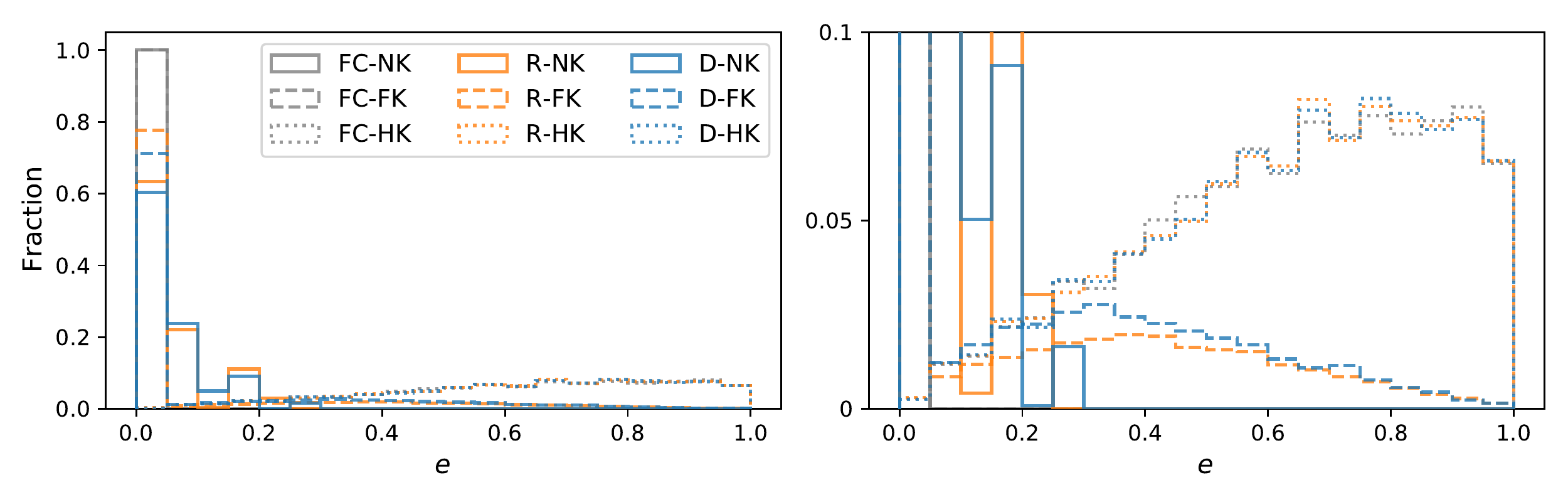}
    \caption{Eccentricity distributions of the identified OB+BH systems for different BH-formation mechanisms. The right panel shows a y-axis zoom-in of the left panel.}
    \label{fig_eccentricities}
\end{figure*}
\begin{table*}
\centering
\caption{Fraction of surviving and identified systems for different BH-formation scenarios. }
\begin{tabu}{ ccc|ccc|ccc }
     \hline
     \hline
     FC- & surviving & identified & R- & surviving & identified &D- & surviving& identified\\ 
     \hline
     \multirow{2}{*}{NK} & \multirow{2}{*}{100\%} & \multirow{2}{*}{68\%} & NK & 88\% & 57\%  & NK & 100\% & 43\%\\ 
      && & FK & 70\% & 58\% &  FK & 61 \%& 57\%\\
     HK & 15\% & 40\%& HK & 13\%& 37\% & HK & 13 \%& 34\%\\
     \hline
\end{tabu}
         \flushleft
    \begin{tablenotes}
      \small
      \item \textbf{Notes.} The fraction of identified systems are systems with $6<G<20$ that are detected as binaries by \Gaias and which are identifiable in the AMRF diagram.
    \end{tablenotes}

\label{table_different_kicks}
\end{table*}

\begin{figure}
    \centering
    \includegraphics[width = 0.49\textwidth]{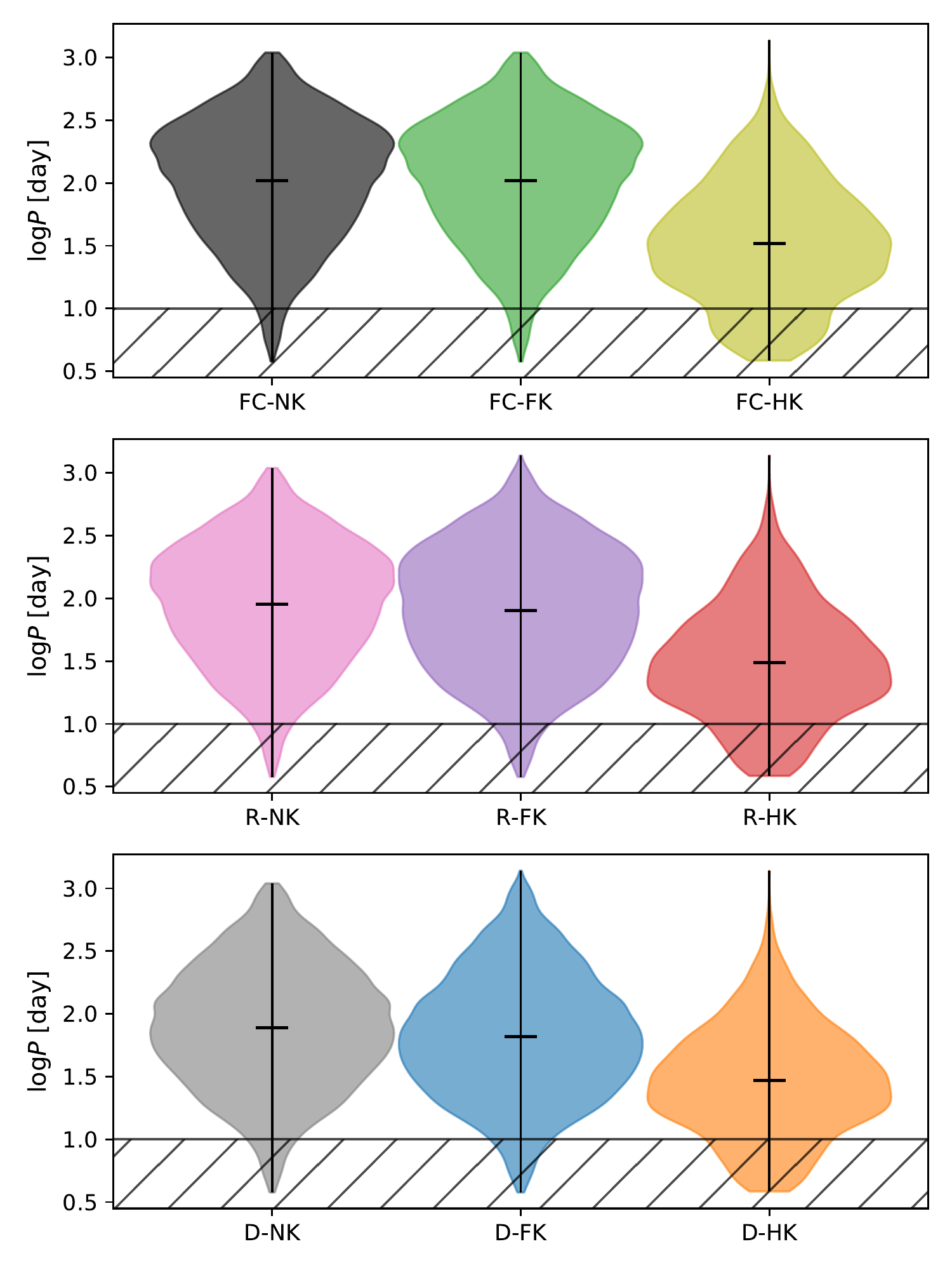}
    \caption{Violin plots for the periods of different BH-formation scenarios, indicated on the $x$-axis. The filled regions are (vertical) probability density plots, which are mirrored over the vertical solid black line. The mean of the periods is indicated with a black horizontal bar. Hatched regions corresponds to possible wind-fed or Roche-lobe overfilling X-ray binaries.}
    \label{fig_periods}
\end{figure}

\begin{figure}
    \centering
    \includegraphics[width = 0.49\textwidth]{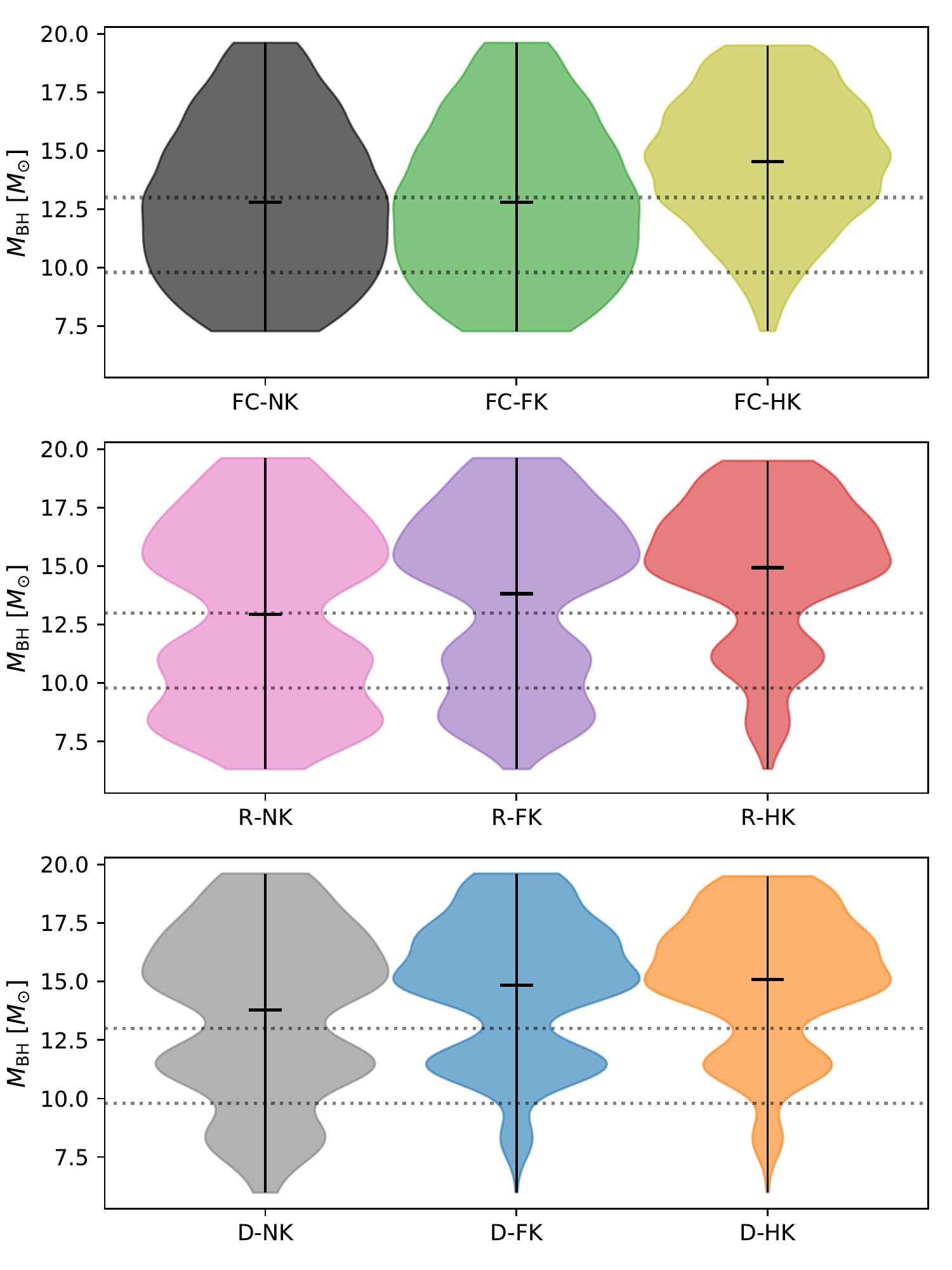}
    \caption{Same as \Figref{fig_periods} but for the masses of the BHs. The distributions shown here are smoothened. The dashed lines correspond to masses around 10 and 13\,$\Modot$.}
    \label{fig_BH_masses_indiv}
\end{figure}

\begin{figure}
    \centering
    \includegraphics[width = 0.49\textwidth]{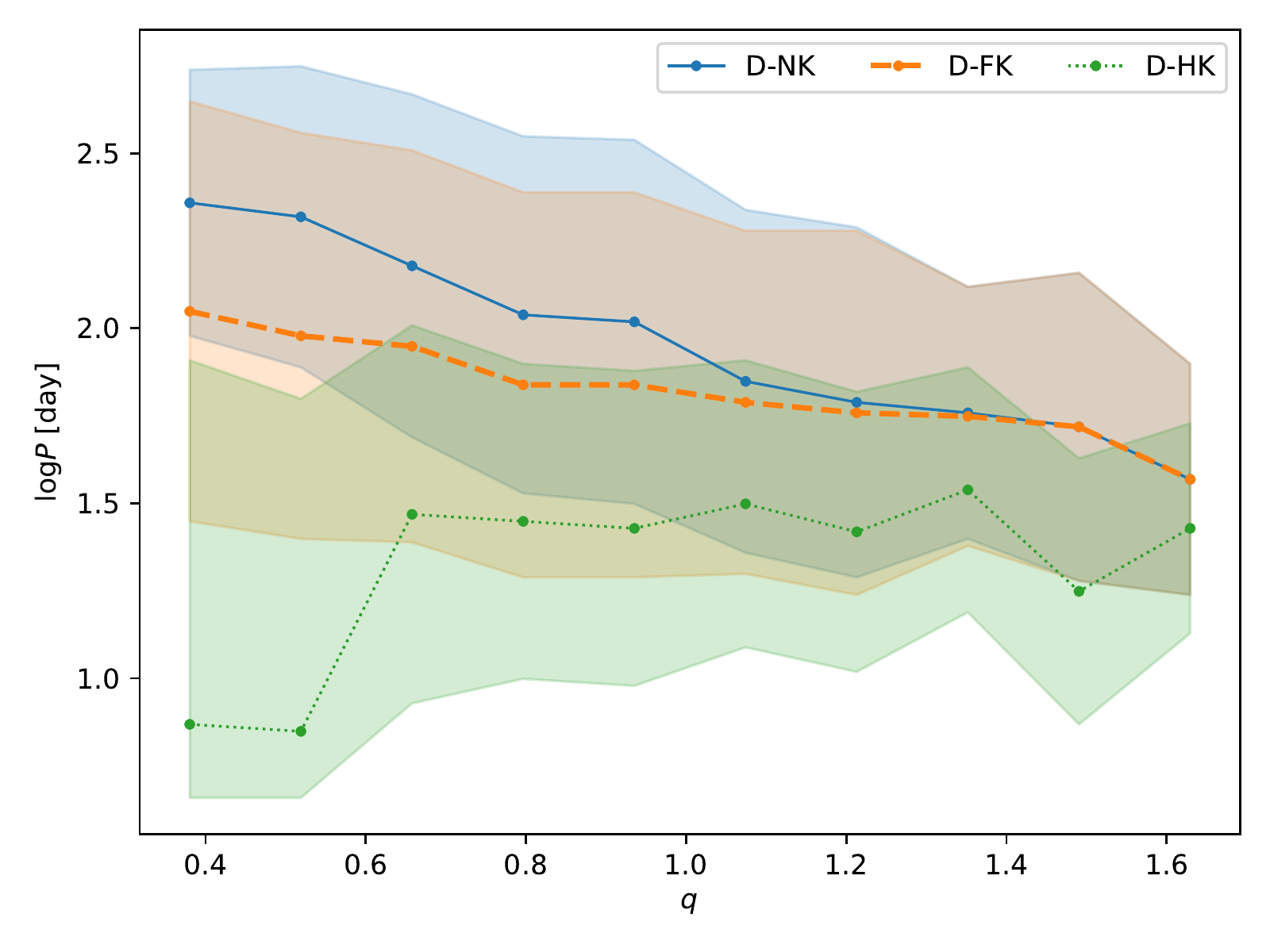}
    \caption{Correlation between mass ratio $q$ and period $P$ of the identified OB+BH systems for the three different kick scenarios combined with the delayed supernova mechanism. Lines represent the modes of the 2D distributions in \Figref{fig_P_vs_q_2dhist} and the shaded regions account for 68\% of the systems.}
    \label{fig_P_vs_q_1dhist}
\end{figure}

The following results are only for DR3. The observed eccentricity distribution of the identified post-SN OB+BH systems is shown in \Figref{fig_eccentricities}, for the aforementioned BH formation cases. If BHs are formed with NS kicks, more systems should be found with larger eccentricities than with low eccentricties, whereas for the fallback kick it is more evenly distributed. In case of no kick, we expect no eccentricities larger than 0.4. \nextline
Different kick mechanisms also result in different observed period distributions. Period distributions for the different BH-formation scenarios are shown in \Figref{fig_periods}. In case of no kick, we expect to observe more long-period systems, whereas for the NS kick scenario more short-period systems and almost no very long-period systems are expected. This is because long-period systems are more easily disrupted by a kick as they are less gravitationally bound than short period systems. The hatched regions in \Figref{fig_periods} indicate the orbital period range ($P\leq 10$\,days) for which wind-fed or Roche-lobe filling X-ray binaries (e.g. Cyg X1) may occur. However, \citet{Sen_2021} predict that even in this region, not many OB+BH binaries will be X-ray bright.\nextline
Different explosion mechanisms lead to distinct distributions of the BH masses of the identified OB+BH systems, shown in \Figref{fig_BH_masses_indiv}. Here, the distributions are smoothened with a Gaussian bandwidth of 0.3 to focus on their overall shape, instead of their discreteness (originating from the discrete distribution of the main-sequence primary star masses in the simulations). We can see that in the rapid (delayed) explosion scenarios, we do not see a continuous distribution as in the case of full collapse, but rather a bimodal (trimodal) one, with a dearth of BHs around 10 (10 and 13)\,$\Modot$ (the dashed lines in \Figref{fig_BH_masses_indiv}) due to the mass lost in the SN. Moreover, for the delayed explosion scenario, we expect BH masses in the mass gap below 5\,$\Modot$, which is not expected for the other two scenarios. Finally, in the case of a NS kick, all scenarios tend towards larger BH masses. 
\nextline
Furthermore, the correlation between the period and the mass ratio can also provide information on the kick scenario, as \Figref{fig_P_vs_q_1dhist} shows. The lines in the figure show the modes of the columns in the 2D distributions in the bottom panels of \Figref{fig_P_vs_q_2dhist} and the shaded regions represent the period boundaries in which 68\% of the systems fall. The difference between the NK and FK scenarios is subtle. For the NK scenario, we expect for the low-mass ratio-systems more detections with longer periods ($P>100$\,days) than shorter periods, gradually lowering the most expected periods as the mass ratio increases. The FK scenario seems to favor a slightly flatter distribution. In the HK scenario, longer periods appear to be less common, as expected and already seen from \Figref{fig_periods}. This is because longer period systems are more easily disrupted by kicks as they are less gravitationally bound. A similar reasoning can be made for the correlation between the mass of the LC and the period. Here, the relative amount of long-period systems amongst the lower-mass LCs rapidly decreases between the NK, FK, and HK scenarios (see \Figref{fig_P_vs_mlc_2dhist}). \nextline
This experiment shows that we should be able to distinguish between different BH-formation scenarios from the distributions of the eccentricity, period, and BH mass. However, there are certain uncertainties embedded in these distributions. Therefore, if the distributions of the observed parameters follow the predicted distributions of one of the BH-formation scenarios mentioned here, we need to carefully take into consideration any uncertainties possibly affecting the predicted distributions. For example, it is sometimes argued that stars more massive than 30$\Modot$ do not expand much during their evolution, such that mass transfer will never be initiated before BH formation of the primary \citep[][]{Gilkis_2021}. Hence, the system will not be circularised by mass transfer. For wider systems in which tides do not play a significant role, we will for example observe the initial eccentricity in the direct-collapse scenario. If this is indeed the case, we might observe large eccentricities which are not related to the BH-formation scenario and originate in the natal eccentricity distribution.
\subsection{Uncertainties on the results}\label{sec_uncertainties}
The analysis performed in this paper has not taken into account uncertainties on the observed parameters. However, observational uncertainties on $\alpha$, $\varpi$, $M_1$, $P$, and the apparent $G$-band magnitude result in uncertainties on the binary detection fraction from \Gaias and the identification fraction in the AMRF diagram. There are two types of contributions to the uncertainties: true-negatives and false-positives. The first one are systems which are OB+BH systems, but which are not detected by our methods, whereas the latter are OB+MS systems that are falsely classified as OB+BH systems in the AMRF diagram. Although, theoretically OB+MS systems all lie below the maximum of the main-sequence-AMRF curve, observational uncertainties could cause some OB+MS systems to falsely pop up above the triple-AMRF. Below, we investigate the effect of both contributions through a Monte Carlo simulation and show that the results presented in \Secref{sec_BH_distributions_title} are robust.

\subsubsection{True negatives}
The estimated number of OB+BH systems in the ALS\,II is around 300. Hence, we draw a sample of 300 OB+BH systems from the simulations and obtain parameters according to the methods explained in \Secref{sec_Gaia_cuts_title}.
\nextline 
For the Monte Carlo simulation, 1000 samples were simulated with new observed parameters, drawn from a normal distribution around the true value with a sigma corresponding to the uncertainty on the parameter. The parameter with the largest uncertainty will most likely be the mass of the primary, since there is no visible binary companion to accurately determine the primary mass from. Therefore, a conservative uncertainty of 50\% was assumed for the mass of the LC, with a lower limit of $7\Modot$. The uncertainties on $\alpha$ and $\varpi$ were both determined by the \Gaias precision at the true magnitude, where we assume \Gaias will always return $\alpha>0$ but also that $\varpi >0$ since the ALS\,II systems have high quality data. A standard deviation of 10\% was used for the period, with a lower limit of 0 days. The uncertainty on $G$ is taken to be 0.01mag \citep[][]{Kourniotis_2014}.\nextline
The detection and identification of the new systems as binaries follows the same observational constraints as explained in \Secref{sec_Gaia_cuts_title} and \Secref{sec_shahaf_title}, respectively. Using again the conservative criterion, the average fraction of OB+BH systems detected by \Gaias is 77.4$^{+2.5}_{-2.1}$\%, where the errors represent a 99.7\% (3-$\sigma$) confidence interval.
The average AMRF completeness is 82.9$^{+4.7}_{-5.5}$\% within a 99.7\% confidence interval. Ultimately, we find that we could identify 64.2$^{+3.5}_{-3.9}$\% of the OB+BH systems that are observed by \Gaia, again within a 99.7\% confidence interval. Although the AMRF completeness is smaller than the one obtained in \Secref{sec_BH_distributions_title}, the total fraction of identified systems is still compatible with the one obtained in \Secref{sec_BH_distributions_title}. Repeating this experiment multiple times with a different sample of 300 OB+BH systems results in similar conclusions. While the detection fraction and AMRF completeness may vary, the overall identification fraction amongst the observed sources are compatible with one another.


\subsubsection{False positives}
Similarly, we investigated how many OB+MS systems would have an observational $\mathcal{A}$ larger than the maximum of the triple-AMRF. Assuming continuous star formation and an intial binary fraction of 70\%, approximately 50\% of massive stars are currently in binaries with a main-sequence companion, while 20\% are post-interaction binaries, and 30\% are single \citep[][]{De_Mink_2014}. This results in about 7000 OB+MS systems present in the ALS\,II. These 7000 OB+MS systems are drawn from the distributions determined by \citet{Sana_2012} and a second set of initial systems is drawn from the distributions presented in \citet{Moe_2017}. In both scenarios, primary masses follow the \citet{Salpeter_1995} initial mass function, ranging from 8 to 60\,$\Modot$.\nextline
Performing multiple monte carlo simulations using the same uncertainties as for the OB+BH systems, we find that, at most, 4 such OB+MS systems will have an observational $\mathcal{A}$ larger than the maximum of the triple-AMRF. However, taking into account the uncertainty on $\mathcal{A}$, none of these systems end up having $\mathcal{A}-\sigma_{\mathcal{A}}>\mathcal{A}_{\mathrm{triple}}$, such that, in principle, they would not be classified as OB+BH systems. Hence, the contribution of OB+MS to the number of false positives to the sample should be negligible.\nextline
In principle, one would want to check the contribution of triples to the number of false positives. However, in reality, the distributions of their parameters are not well known \citep[see e.g.][]{Duchene_2013}. Performing such analysis is hence beyond the scope of this research. However, it should be taken into account for further analysis.

\subsection{OB+BH systems in the extended mission}
We can also investigate how the detection of OB+BH systems (in the FC-NK scenario) would improve in the extended 10-yr mission, which is already ongoing with a mid-term review in 2022\footnote{\url{https://sci.esa.int/web/gaia/-/47354-fact-sheet}}. Assuming that the precision at the end of the extended 10-yr mission will not be limited by instrumentation and that the number of observations doubles with respect to the 5-yr mission, we can scale the precision with a factor $\sqrt{5}/\sqrt{10}$. The conservative upper limit on the period increases to 10\,yr.\nextline
The fraction of systems detected by \Gaias slightly increases to 88.3\% of the sampled OB+BH systems. The AMRF completeness is still fairly high with a value of 80.0\%, yielding a conservative identification of 70.7\% of the sampled OB+BH systems), which is 1\% more than at the end of the 5-yr mission. In terms of true numbers, this would mean an identification of $\sim$200 OB+BH binaries of the sources in ALS\,II, not including the bright sources. An overview of the distributions of the identified OB+BH binaries is given in \Figref{fig_observed_distributions_10yr}. As a reference, the distributions of the identified systems in DR3 and DR4 are also shown. While the contribution of the extended mission is small in absolute numbers, it will allow us to probe the longer end of the period distribution ($P > 5$\,yr) where critical signatures of the kick scenario are expected (see \Figref{fig_periods}).\nextline
The number of sources in the \Gaias catalogue has increased over different data releases. On the one hand there is the better precision with which sources can be observed. On the other hand, for very faint and bright sources, more data also means a (better) detection. On top of that, the known catalogue of massive stars will increase over time (e.g. ALS\,III). Therefore, the estimated number might even be much higher at the end of the extended mission, as even more sources could be observed by \Gaias and uncertainties on the observed parameters could become smaller.

\section{Summary}\label{sec_summary}
We have used massive star evolution computations of \citet{Langer_2020}, which assume a direct collapse and no natal kick upon BH formation, as fiducial to study the expected distribution of OB+BH binaries, i.e. the BH\,+\,BH/NS progenitors, in the Milky Way. In particular, we have investigated how many OB+BH systems that are observed by \Gaia, will be detected as binaries, both in DR3 and in the DR4 catalogue at the end of the 5-yr mission. Using the distance distribution of known OB stars in the ALS\,II catalogue, we find that, in DR3, 76.5\% of the OB+BH systems observed by \Gaias will be classified as binaries in the \Gaias catalogue in the conservative case. In DR4, this number increases to 85.0\%. These fractions can change when using a significantly different distance distribution.\nextline
We have also adapted the method established by \citet{Shahaf_2019} for massive primary components, to identify the detected OB+BH binaries. We can identify 88.7\% or 82.3\% of the OB+BH binaries that are detected by \Gaias as binaries in DR3 or DR4, respectively. This ultimately leads to the identification of 67.8\% of the OB+BH binaries observed by \Gaias in DR3, while 69.9\% of the observed OB+BH binaries can be identified in DR4. Translating these fractions to an absolute number of OB+BH systems in the ALS II, around 190 OB+BH binaries can be identified as such already in DR3, increasing the known (quiescent) OB+BH binaries by a factor of more than 20. An additional $\sim$5 OB+BH systems are identifiable in DR4, mostly with short ($P \lesssim 10\,$d) or long periods ($P \gtrsim 1000\,$d).\nextline
We also found that different BH-formation mechanisms result not only in different detection fractions, but also in distinct distributions of BH masses, orbital periods, and eccentricities. However, relating the distributions of the identified OB+BH systems directly to the BH-formation mechanism should be done with caution as different scenarios may occur for different mass regimes. Nonetheless, our results predict that both the numbers and the distributions of orbital properties of the detected OB+BH systems will provide crucial observational constraints to the collapse scenario and the formation of BH+BH/NS progenitors.\nextline
We have shown that the identifiable fraction of OB+BH systems is robust. Hence, the number of OB+BH system identifications opens up the possibility to draw profound conclusions on the physics of BH formation. 

\begin{acknowledgements}
SJ acknowledges support from the FWO PhD fellowship under project 11E1721N. 

This project has received funding from the European Research Council (ERC) under the European Union's Horizon 2020 research and innovation programme (grant agreement n$\degree$ 772225/MULTIPLES).

This research was supported by Grant No.~2016069 of the United States-Israel Binational Science Foundation (BSF), and by Grant
No. I-1498-303.7/2019 of the German-Israeli Foundation for Scientific Research and Development (GIF).
\end{acknowledgements}

\bibliography{References}

\newpage
\begin{appendices}
\renewcommand{\thefigure}{A\arabic{figure}}
\setcounter{figure}{0}
\renewcommand{\thetable}{A\arabic{table}}
\setcounter{table}{0}

\section{The period change due to mass loss}\label{app_period_change}
We derive here an expression for the final period of a binary system, where the change in period is only caused by mass-loss through winds of the components. The angular momentum loss of a non-eccentric system ($e = 0$, $\dot{e} = 0$) is given by 
\begin{equation}\label{eq_ap_angular_momentum_change}
    3\frac{\dot{J}}{J} = 3\frac{\dot{M}_1}{M_1} + 3\frac{\dot{M}_2}{M_2} - \frac{\dot{M}_{\text{tot}}}{M_{\text{tot}}} + \frac{\dot{P}}{P},
\end{equation}
where for notation simplicity we have used $M_{\text{tot}} = M_1 + M_2$ and $\dot{M}_{\text{tot}} = \dot{M}_1 + \dot{M}_2$.
The specific orbital angular momentum is given by
\begin{equation}
    h_i = \frac{J_i}{M_i} = J_{\text{orb}} \frac{M_{j}}{M_i M_{\text{tot}}},
\end{equation}
where $i$ and $j$ are the components of the binary system. This expression assumes that the radius of component $i$ is much smaller than its Roche-lobe radius, such that spin-angular momentum can be neglected.\nextline
The angular momentum loss can hence be written as $\dot{J} = \dot{J}_1 + \dot{J}_2 = \sum\limits_{1,2} h_i \dot{M}_i$. 
Using this expression in \Eqref{eq_ap_angular_momentum_change} gives
\begin{equation} \label{eq_ap_period_change_inter}
    \begin{split}
       \frac{\dot{P}}{P} = 3\left(\frac{M_2\dot{M}_1}{M_1 M_{\text{tot}}} +  \frac{M_1\dot{M}_2}{M_2 M_{\text{tot}}}\right) - 3\frac{\dot{M}_1}{M_1} - 3\frac{\dot{M}_2}{M_2} + \frac{\dot{M}_{\text{tot}}}{M_{\text{tot}}}.
    \end{split}
\end{equation}

After some simplification, this becomes
\begin{equation}
    \frac{\dot{P}}{P} = -2 \frac{\dot{M}_{\text{tot}}}{M_{\text{tot}}} = -2 \frac{\dot{M}_1 + \dot{M}_2}{M_1 + M_2}.
\end{equation}
Solving this expression gives us \Eqref{eq_final_period}.

\section{Mass and period differences between the LMC and the Milky Way}\label{app_LMCvsMW}
\begin{figure*}[h]
    \centering
    \begin{subfigure}{0.49\linewidth}
    \includegraphics[width = \textwidth]{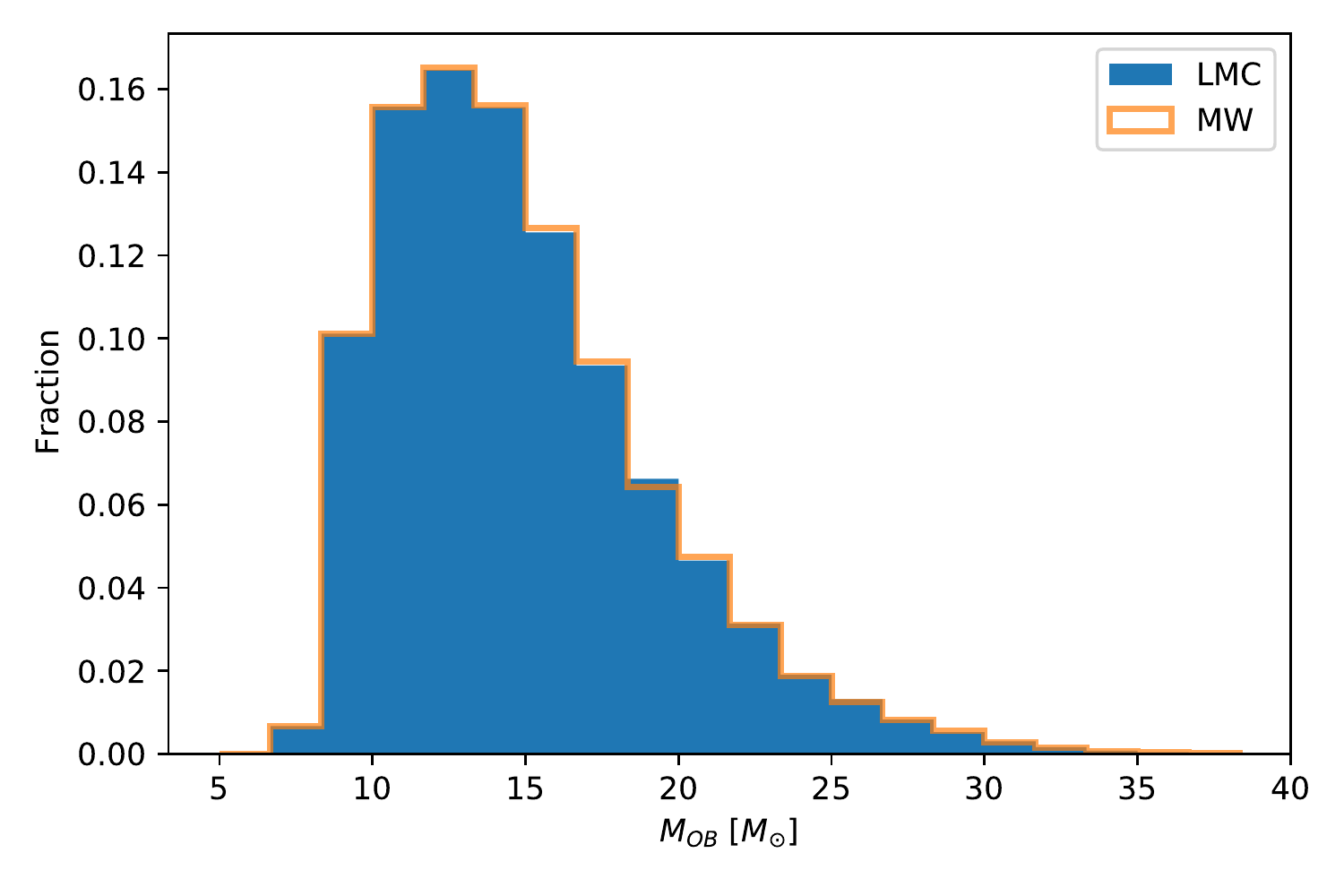}
    \end{subfigure}
    \begin{subfigure}{0.49\linewidth}
    \includegraphics[width = \textwidth]{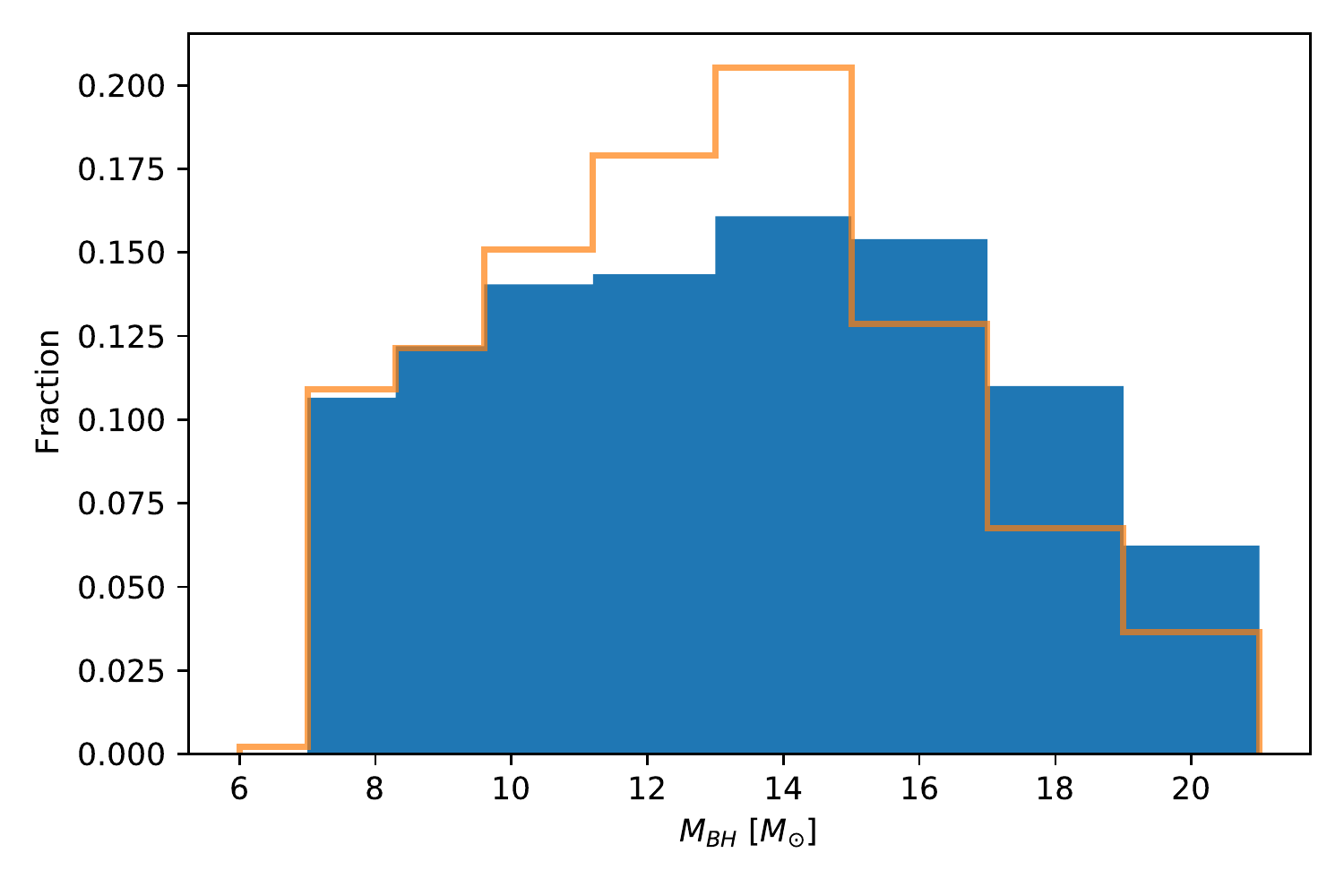}
    \end{subfigure}
    \begin{subfigure}{0.49\linewidth}
    \includegraphics[width = \textwidth]{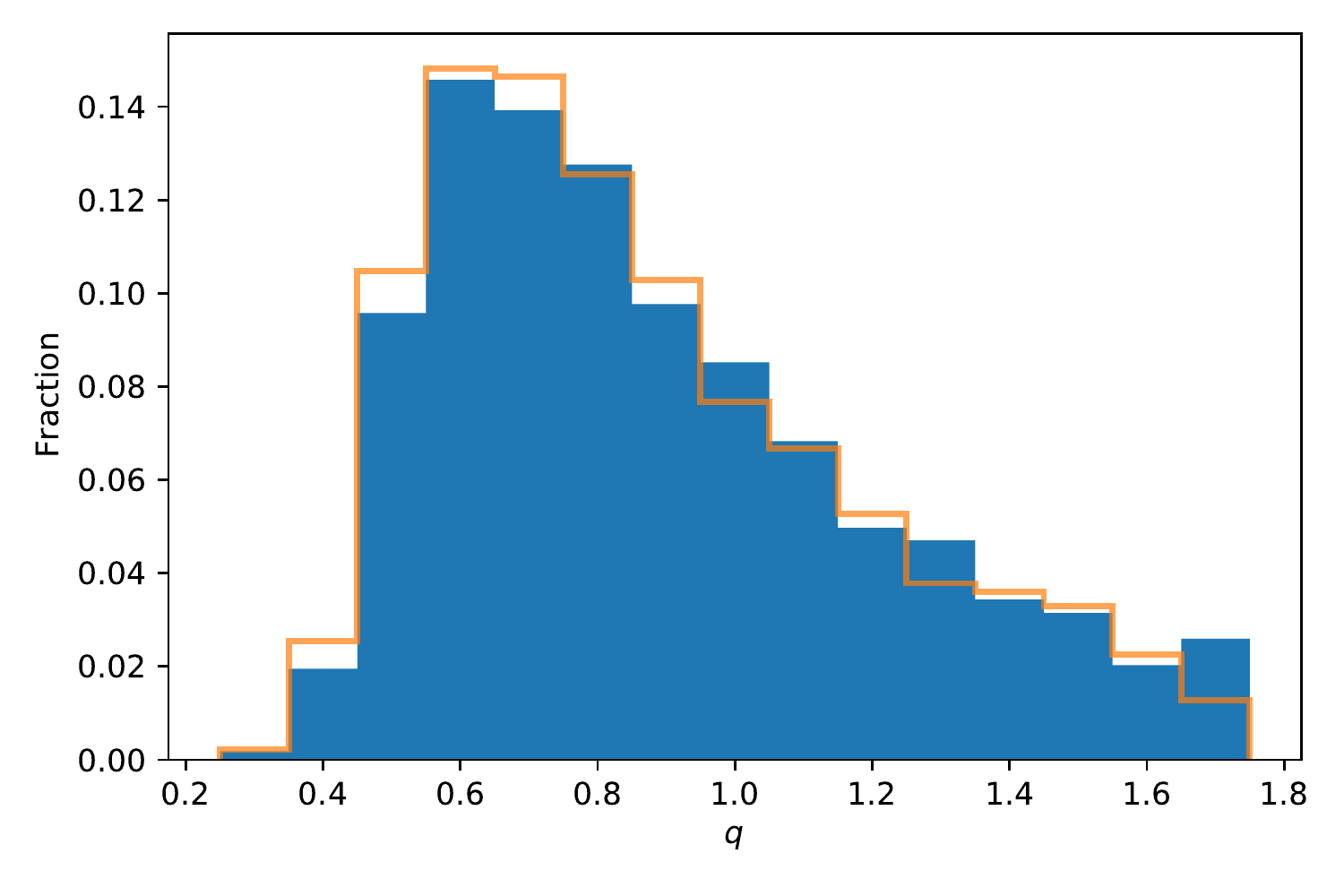}
    \end{subfigure}
    \begin{subfigure}{0.49\linewidth}
    \includegraphics[width = \textwidth]{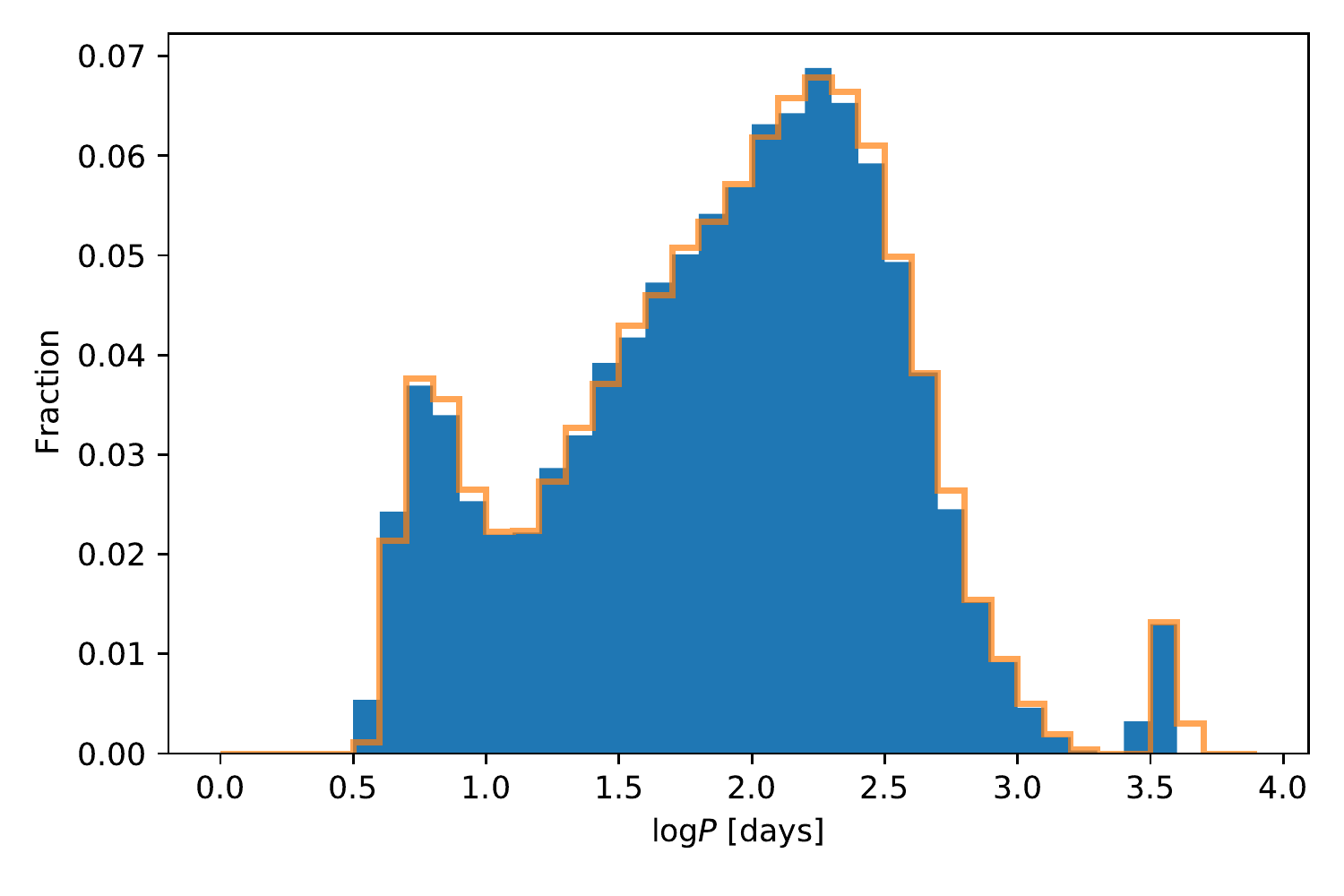}
    \end{subfigure}
    \caption{Difference in distributions of the LC (top-left) and BH (top-right) mass, the mass-ratio ($M_{\text{BH}}/M_{\text{OB}}$, bottom-left), and the period (bottom-right) of the OB+BH systems in the LMC (filled) and Milky Way (open) after applying the corrections mentioned in \Secref{sec_sims_LMC_to_MW}. The mass distributions of the LC are almost identical and minor changes occur due to the relatively weak winds of the main-sequence LC, whereas the masses of the BHs are shifted towards lower values due to the powerful stellar winds experienced by the post-mass-transfer components. Hence, the mass ratios have shifted towards lower values as well. The orbits have slightly widened.}
    \label{fig_LMCvsMW}
\end{figure*}

Figure \ref{fig_LMCvsMW} presents the distributions of the masses of the LCs and BHs, the mass ratios, and periods of the OB+BH systems in the LMC compared to those in the Milky Way.

\newpage~\newpage
\section{The semi-major axis of the projected orbit}\label{app_semi-major_axis_proj}
We derive an expression for the semi-major axis of the projected orbit, starting from the general equation of an ellipse. The general form of an ellipse whose semi-major ($a$) and semi-minor ($b$) axis are not aligned with the axes of the reference frame is given by
\begin{equation} \label{eq_general_ellipse}
    \mathcal{A} x^2 + \mathcal{B}xy + \mathcal{C}y^2 + \mathcal{D}x + \mathcal{E}y+ \mathcal{F} = 0,
\end{equation}
with
\begin{equation} \label{eq_general_ellipse_constants}
    \begin{split}
        &\mathcal{A} = a^2 \sin^2\theta + b^2 \cos^2\theta, \\
        &\mathcal{B} = 2\left( b^2 - a^2 \right)\sin\theta\cos\theta,\\
        &\mathcal{C} = a^2 \cos^2\theta + b^2 \sin^2\theta,\\
        &\mathcal{D} = -2\mathcal{A}x_c - \mathcal{B}y_c,\\
        &\mathcal{E} = -\mathcal{B}x_c - 2\mathcal{C}y_c,\\
        &\mathcal{F} = \mathcal{A}x_c^2 + \mathcal{B}x_c y_c + \mathcal{C}y_c^2 - a^2b^2, \\
        & \mathcal{B}^2 - 4\mathcal{AC} < 0,
    \end{split}
\end{equation}
where $x_c$ and $y_c$ are the coordinates of the centre and $\theta$ is the angle between $a$ and the horizontal axis of the reference frame. For an ellipse centred at the origin, $\mathcal{D},\mathcal{E} = 0$ and $\mathcal{F} = - a^2b^2 = \mathcal{B}^2 /4 -\mathcal{AC}$.
The semi-major axis of the ellipse is given by
\begin{equation} \label{eq_general_ellipse_a}
    a^2 = \frac{\mathcal{A}+\mathcal{C}+\sqrt{(\mathcal{A}-\mathcal{C})^2+\mathcal{B}^2}}{2}.
\end{equation}
\nextline
We can take three points on an ellipse centred at the origin with coordinates ($x_1, y_1$), ($x_2, y_2$), and ($x_3, y_3$). By using \Eqref{eq_general_ellipse} for each of the three points and solving the three equations, the constants $\mathcal{A}$, $\mathcal{B}$, and $\mathcal{C}$ can be written as 
\begin{equation} \label{eq_general_ellipse_ABC}
\begin{split}
      &\mathcal{A} = \mathcal{B}w,\\
      &\mathcal{B} = 4 \frac{w x_1^2 + x_1 y_1 + v y_1^2}{4 vw - 1},\\
      &\mathcal{C} = \mathcal{B} v,
\end{split}
\end{equation}
with 
\begin{equation} \label{eq_general_ellipse_vw}
\begin{split}
    &v = \frac{(x_2y_2-x_1y_1)(x_3^2-x_2^2) + (x_3y_3-x_2y_2)(x_1^2 - x_2^2)}{(y_2^2-y_1^2)(x_2^2-x_3^2) + (y_2^2-y_3^2)(x_1^2-x_2^2)},\\
    &w = \frac{x_2y_2-x_1y_1}{x_1^2-x_2^2} + \frac{y_2^2 - y_1^2}{x_1^2 - x_2^2}v.
\end{split}
\end{equation}
\nextline
We want to obtain an expression for the semi-major axis of the projected ellipse as a function of the orbital parameters $i$, $\Omega$, and $\omega$, independent of the three points chosen. The central coordinates of the projected ellipse, using \Eqref{eq_projected_coordinates}, are $x_c = -Ae$ and $y_c = -Be$. We can transform the coordinates of the projected ellipse to $x - x_c$ and $y - y_c$, such that we implicitly have an ellipse centred at the origin. By doing this, we can use \Eqref{eq_general_ellipse} for a general ellipse centred at the origin. Moreover, the projected coordinates simplify to 
\begin{equation}
    \begin{split}
        x_{\text{proj}} &= A\cos E + F\sqrt{1-e^2}\sin E,\\
        y_{\text{proj}} &= B\cos E + G\sqrt{1-e^2}\sin E.
    \end{split}
\end{equation}
\nextline
Let us start by investigating the variable $v$ in \Eqref{eq_general_ellipse_vw}. The \textit{Thiele-Innes} constants $A$, $B$, $F$, and $G$ (\Eqref{eq_Thiele-Innes}) of any coordinate on the ellipse are equal to each other. Hence, we want to show that $v$ is independent of the eccentric anomaly $E$, as this is the only parameter that changes between different points on the ellipse. One way of verifying this, is by showing that the partial derivative of $v$ with respect to all $E$'s is equals to zero such that for each point $i$ we have $\partial_{E_i}(v) = 0$. If we call $n$ the numerator and $d$ the denominator, this is equivalent to $\partial_{E_i}(n)/\partial_{E_i}(d) = n/d$. By calculating both $d\partial_{E_i}(n)$ and $n\partial_{E_i}(d)$ of $v$, it can be shown that $v$ is independent of $E_1$, $E_2$, and $E_3$. We can hence take $E_1 = 0$, $E_2 = \pi/3$, and $E_3 = 2\pi/3$ to find a simple expression for $v$ as a function of $A$, $B$, $F$, and $G$,
\begin{equation}\label{eq_app_v_simple}
    v = -\frac{1}{2}\frac{A^2+F'^2}{AB+F'G'},
\end{equation}
where for simplicity of notation we have $F' = F\sqrt{1-e^2}$ and $G' = G\sqrt{1-e^2}$.\nextline
By following the same method and using the simplified version of $v$, it can be shown that $w$ is also independent of the choice of $E_1$ and $E_2$. We take $E_1 = 0$ and $E_2 = \pi/2$ and simplify $w$ to
\begin{equation}\label{eq_app_w_simple}
    w = -\frac{1}{2}\frac{B^2 + G^2}{AB+F'G'}.
\end{equation}
\nextline
We now only need to show that $\mathcal{B}$ is also constant. The same method can again be used. We use Eqs. (\ref{eq_app_v_simple},\ref{eq_app_w_simple}) and take $E_1 = 0$, such that
\begin{equation}
    \mathcal{B} = - 2(AB+F'G').
\end{equation}
\nextline
The equation for the semi-major axis of the projected ellipse `centred' at the origin is obtained by using the above three expressions in \Eqref{eq_general_ellipse_ABC} and \eqref{eq_general_ellipse_a},
\begin{equation} \label{eq_general_ellipse_a_final}
\begin{split}
    a^2 &= \frac{\mathcal{B}w + \mathcal{B}v + \sqrt{(\mathcal{B}w - \mathcal{B}v)^2 + \mathcal{B}^2}}{2}\\
    &= -(AB+F'G')(w + v) + \left| -(AB+F'G')\right| \sqrt{(w-v)^2 +1},
\end{split}
\end{equation}
where $w$ and $v$ are now constants only dependent on the orbital parameters through the \textit{Thiele-Innes} constants. For circular orbits, $e = 0$ and we have $F' = F$ and $G' = G$. By working out each of the terms in \Eqref{eq_general_ellipse_a_final}, it can be shown that the semi-major axis of the projected ellipse is equal to the true semi-major axis of the circular orbit.

\section{Masses, effective temperatures, and radii}
Table \ref{table_data_mass-mag} shows the literature data obtained for main-sequence dwarf stars.
\begin{table}[!b]
\renewcommand{\arraystretch}{1.1}
    \centering
    \caption{Masses, effective temperatures and radii of dwarf stars used for the fitting of the mass-magnitude relation in \Secref{sec_mass-magnitude_relation}.}
     \begin{tabular}{lllll}
        \hline \hline
        Mass [$\Modot$] & $\Teff$ [K]& $R$ [$\Rodot$]& SpT & Ref.\\
        \hline
       57.95 & 44852 & 13.80 & O3\,V & a\\
       46.94 & 42857 & 12.42 & O4\,V & a\\
       38.08 & 40862 & 11.20 & O5\,V & a\\
34.39 & 39865 & 10.64 & O5.5\,V & a\\
30.98 & 38867 & 10.11 & O6\,V & a\\
28.00 & 37870 & 9.61 & O6.5\,V & a\\
25.29 & 36872 & 9.15 & O7\,V & a\\
22.90 & 35874 & 8.70 & O7.5\,V & a\\
20.76 & 34877 & 8.29 & O8\,V & a\\
18.80 & 33879 & 7.90 & O8.5\,V & a\\
17.08 & 32882 & 7.53 & O9\,V & a\\
15.55 & 31884 & 7.18 & O9.5\,V & a\\
13.21 & 25400 & 6.42 & B1\,V & b\\
9.11 & 20800 & 5.33 & B2\,V & b\\
7.6 & 18800 & 4.8 & B3\,V & b\\
5.90 & 15200 & 3.90 & B5\,V & b\\
5.17 & 13800 & 3.56 & B6\,V & b\\
4.45 & 12400 & 3.28 & B7\,V & b\\
3.80 & 11400 & 3.00 & B8\,V & b\\
3.29 & 10600 & 2.70 & B9\,V & b\\
2.40 & 9727 & 1.87 & A0\,V & c\\
2.19 & 8820 & 1.78 & A2\,V & c\\
1.86 & 7880 & 1.69 & A5\,V & c\\
1.8 & 7672 & 1.66 & A6\,V & c\\
1.74 & 7483 & 1.63 & A7\,V & c\\
1.66 & 7305 & 1.6 & A8\,V & c\\
1.62 & 7112 & 1.55 & A9\,V & c\\
1.55 & 6949 & 1.51 & F0\,V & c\\
1.236 & 6516 & 1.32 & F6\,V & d\\
1.05 & 5943 & 1.07 & G0\,V & e,f\\
1.00 & 5794 & 1.02 & G2\,V & e,f\\
0.95 & 5495 & 0.91 & G5\,V & e,f\\
0.91 & 5248 & 0.83 & G8\,V & e,f\\
0.806 & 5246 & 0.778 & K0\,V & g\\
0.781 & 5077 & 0.735 & K2\,V & g\\
0.81 & 4699 & 0.778 & K3\,V & h\\
0.690 & 4400 & 0.665 & K5\,V & i\\
0.6 & 3800 & 0.62 & M0\,V & j\\
0.49 & 3600 & 0.49 & M1\,V & j\\
0.44 & 3400 & 0.44 & M2\,V & j\\
0.36 & 3250 & 0.39 & M3\,V & j\\
0.20 & 3100 & 0.26 & M4\,V & j\\
        \hline
     \end{tabular}
         \flushleft
    \begin{tablenotes}
      \small
      \item \textbf{Notes.} Data from: $^{\text{(a)}}$ \cite{Martins_2005}; $^{\text{(b)}}$ \citet{Silaj_2014}; $^{\text{(c)}}$ \citet{Adelman_2004}; $^{\text{(d)}}$ \citet{Boyajian_2012}; $^{\text{(e)}}$ \citet{Straizys_1981}; $^{\text{(f)}}$ \citet{Vardavas_2011}; $^{\text{(g)}}$ \citet{Boyajian_2012_II}; $^{\text{(h)}}$ \citet{Gillon_2017}; $^{\text{(i)}}$ \citet{Kervella_2008}; $^{\text{(j)}}$ \citet{Kaltenegger_2009}
    \end{tablenotes}
    \label{table_data_mass-mag}
\end{table}

\section{Corner plots for the detected systems in \Gaia}
\begin{figure*}
    \centering
    \includegraphics[width = 0.98\textwidth]{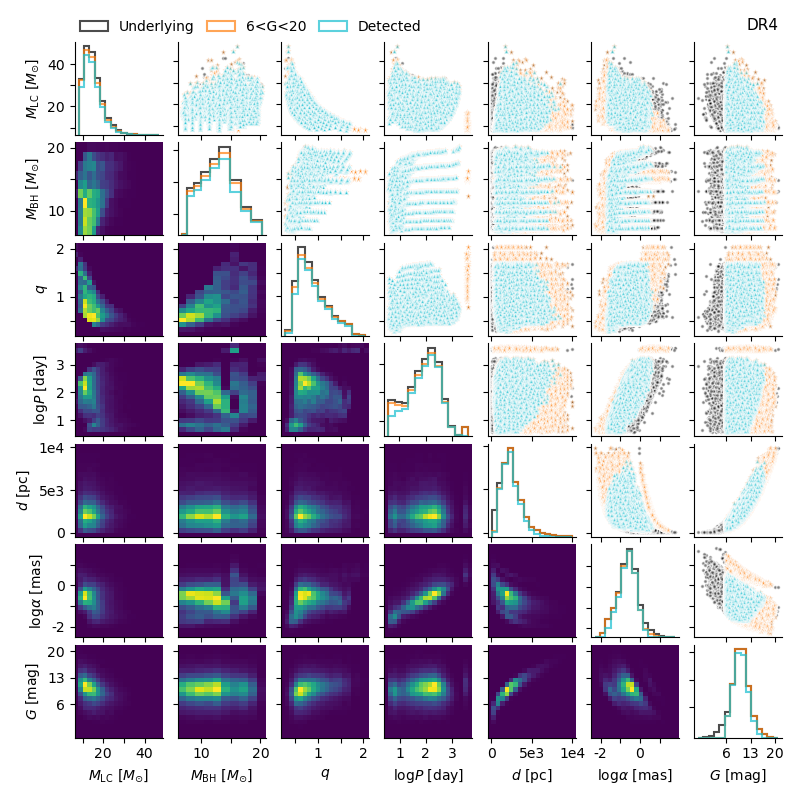}
    \caption{Same as \Figref{fig_corner_3yr} but for DR4.}
    \label{fig_corner_5yr}
\end{figure*}

Figures \ref{fig_corner_3yr} and \ref{fig_corner_5yr} show corner plots for the parameters of OB+BH systems in DR3 and DR4, respectively, after applying several observational constraints.
\noindent%
\begin{minipage*}{\textwidth}
\makebox[\textwidth]{
  \includegraphics[width = \textwidth]{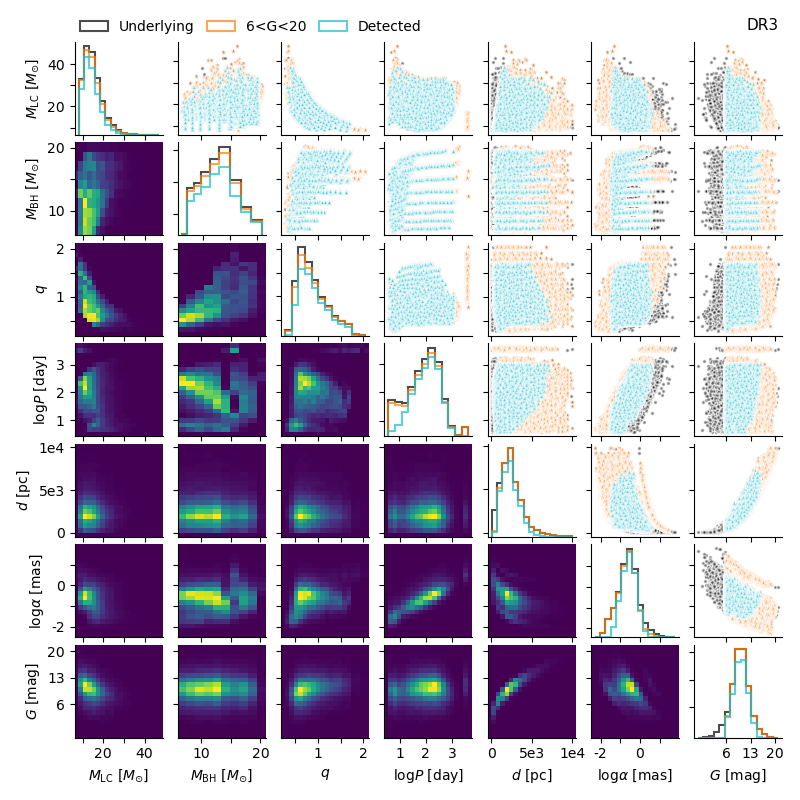}}
    \captionof{figure}{Corner plot for the OB+BH systems in \Gaias in DR3. Black represents the underlying distribution of the simulated systems, orange are the systems which are detected by \Gaias (i.e. have a magnitude $6<G<20$), and cyan are the detected systems in the conservative case. The under-diagonal plots are 2D histograms showing the underlying distribution.}
    \label{fig_corner_3yr}
\end{minipage*}

\newpage~\newpage~\newpage~\newpage

\section{Mass and period distribution for different BH formation mechanisms}\label{app_kicks_velocities}
\begin{figure*}
    \centering
     \includegraphics[width = \textwidth]{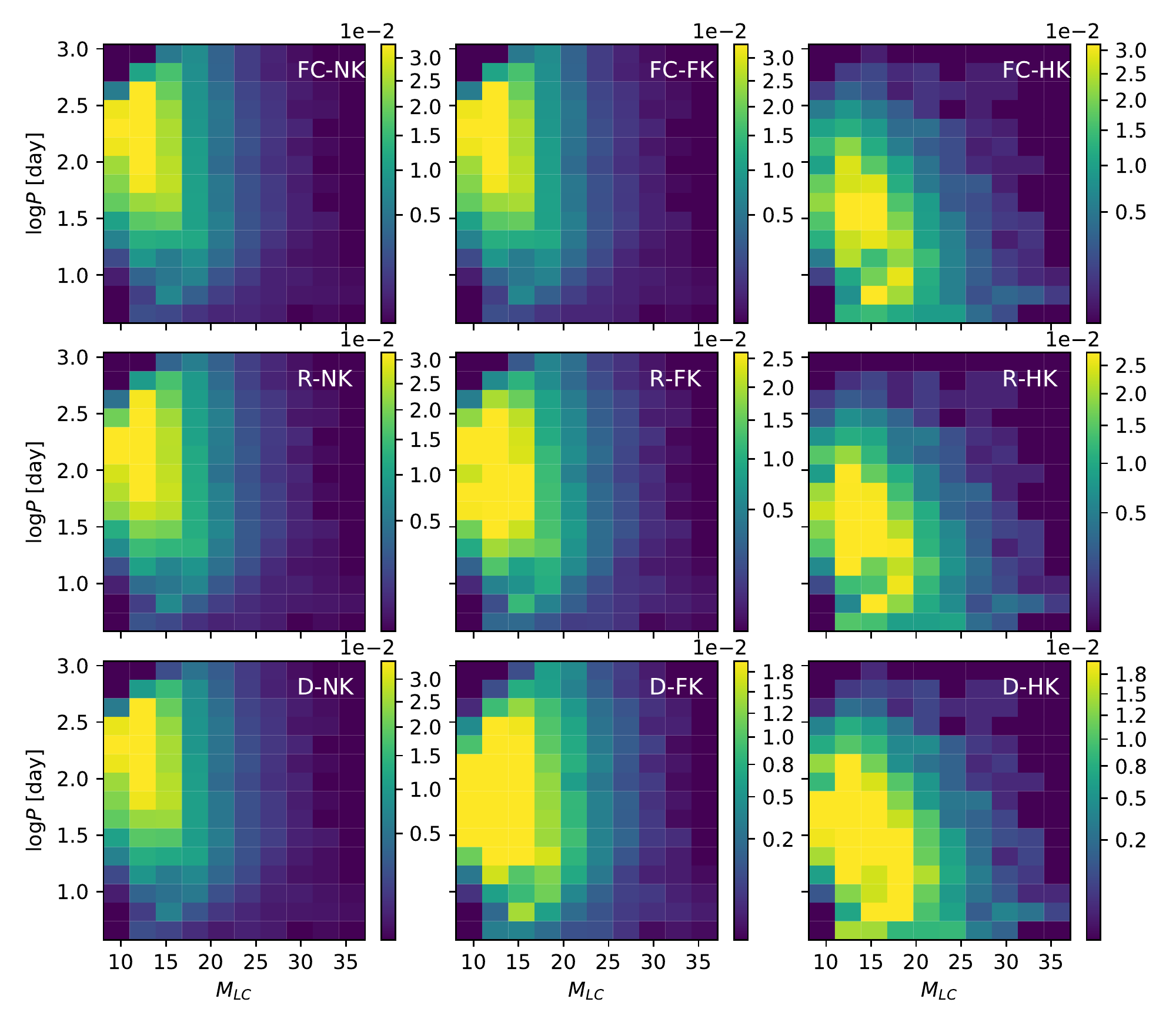}
    \caption{Two-dimensional histograms of the mass of the LC $M_{\text{LC}}$ and the period $P$. The histograms look similar for different collapse scenarios, but not for different kick scenarios. 
    }
    \label{fig_P_vs_mlc_2dhist}
\end{figure*}

Figures \ref{fig_P_vs_q_2dhist} and \ref{fig_P_vs_mlc_2dhist} show the correlation distributions between the period and the mass ratio, and, the period and mass of the LC, respectively, for the different BH-formation scenarios. 
\noindent%
\begin{minipage*}{\textwidth}
\makebox[\textwidth]{
  \includegraphics[width = \textwidth]{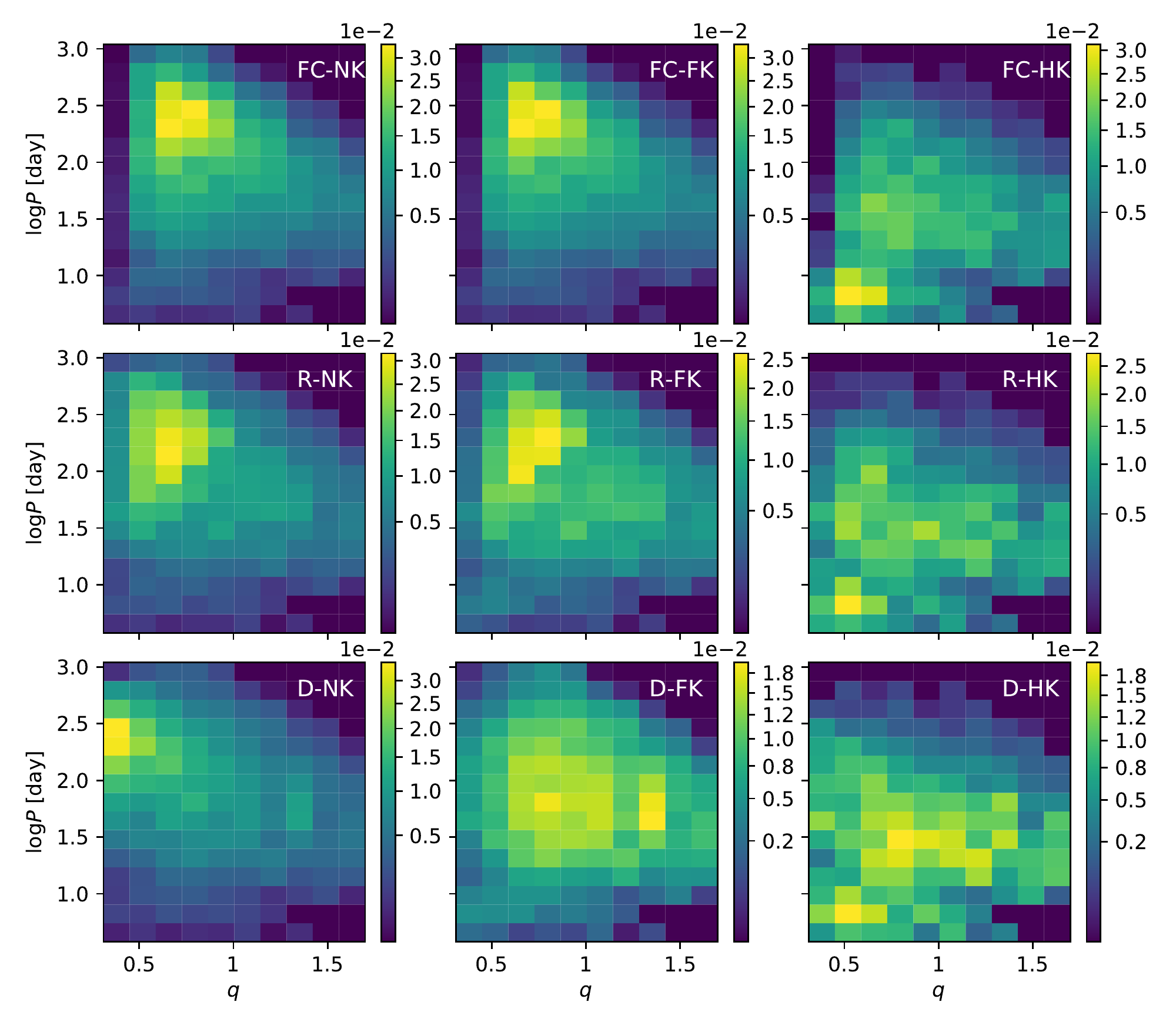}}
    \captionof{figure}{Two-dimensional histograms of the mass ratio $q$ and the period $P$. The histograms look similar for different collapse scenarios, but not for different kick scenarios. }
    \label{fig_P_vs_q_2dhist}
\end{minipage*}

\newpage~\newpage~\newpage
\section{Mass and period distribution in the 10-yr mission}
\begin{figure*}
    \centering
    \begin{subfigure}{0.495\linewidth}
    \includegraphics[width = \textwidth]{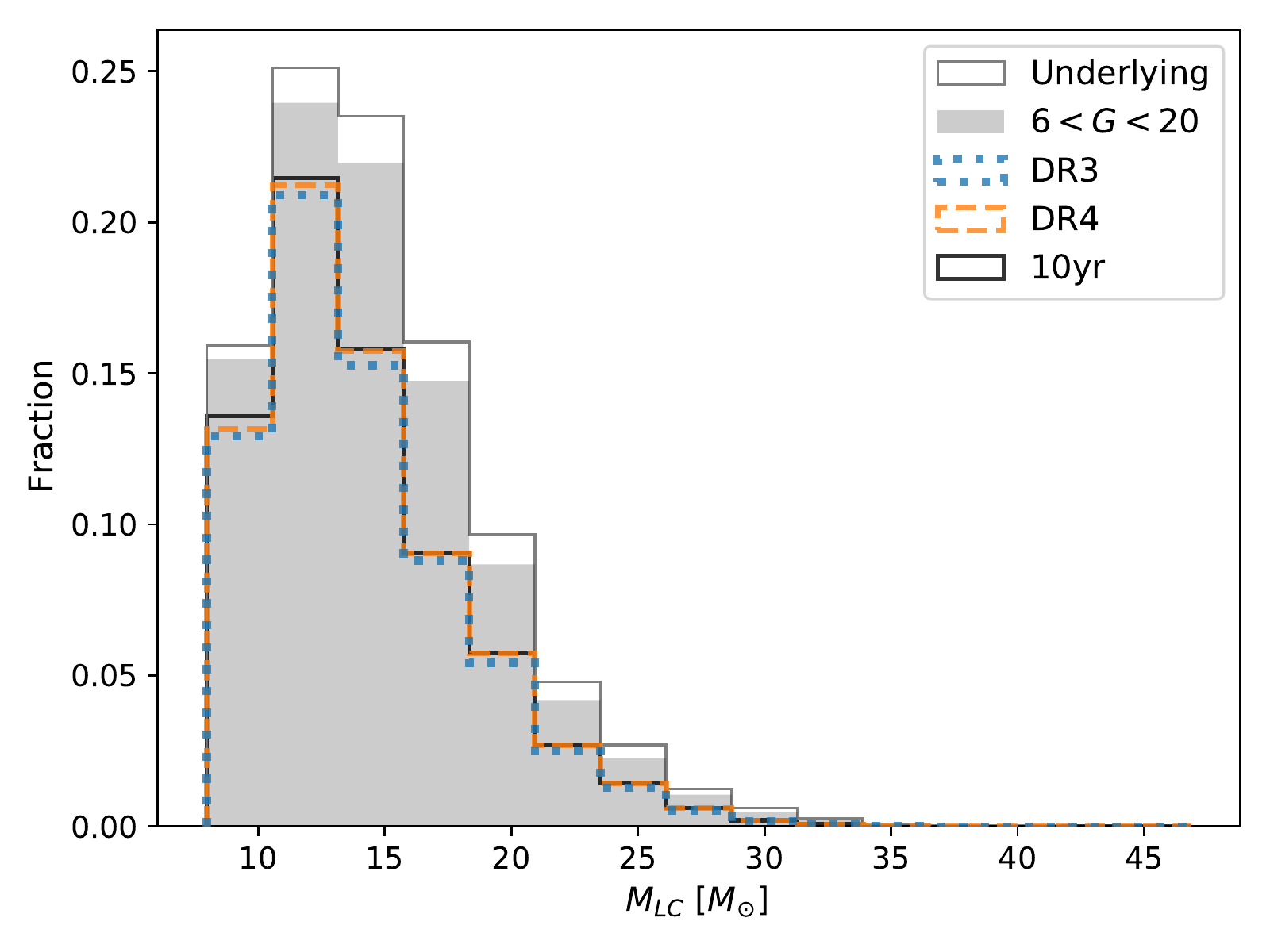}
    \end{subfigure}
    \begin{subfigure}{0.495\linewidth}
    \includegraphics[width = \textwidth]{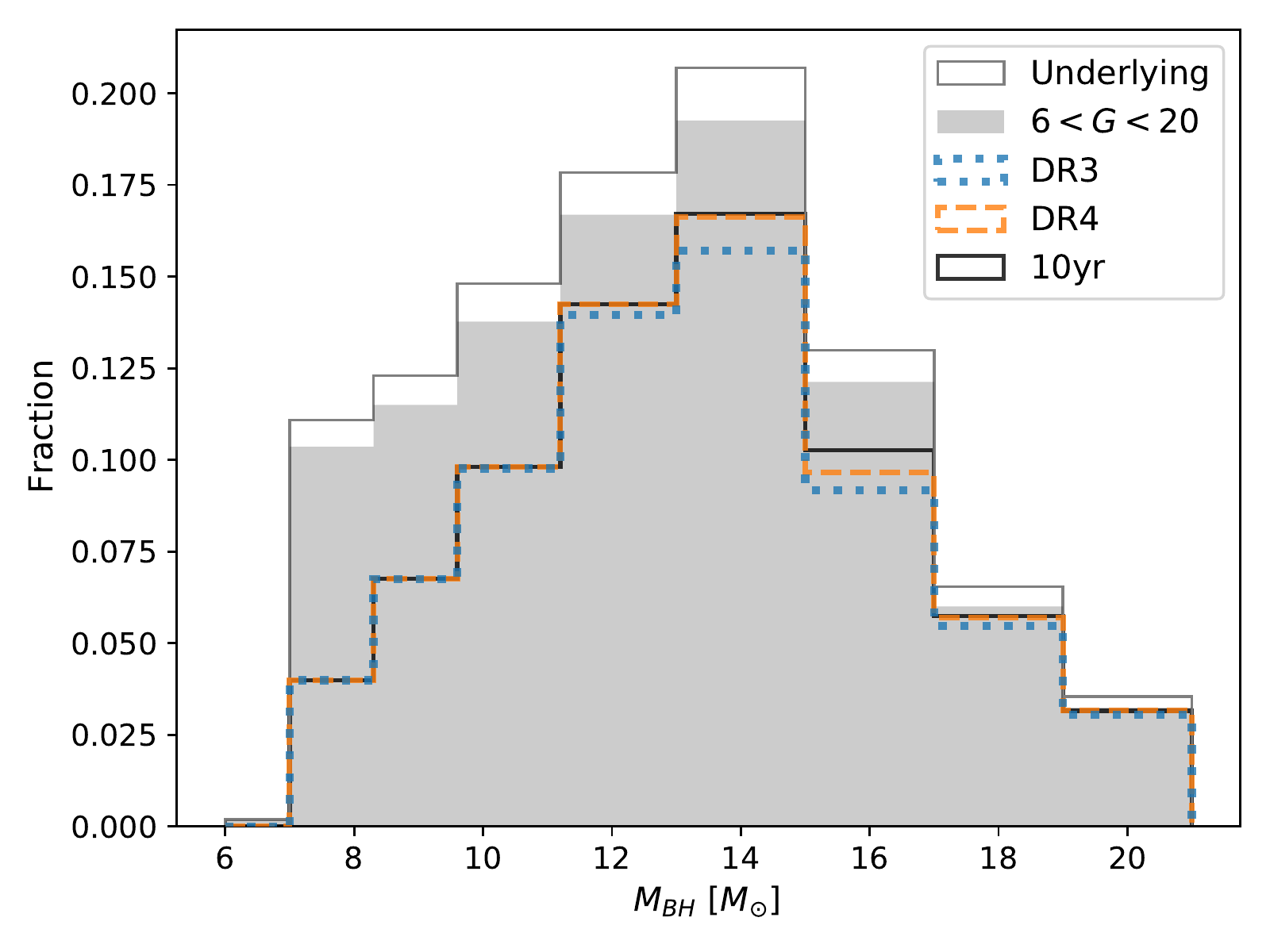}
    \end{subfigure}
    \begin{subfigure}{0.495\linewidth}
    \includegraphics[width = \textwidth]{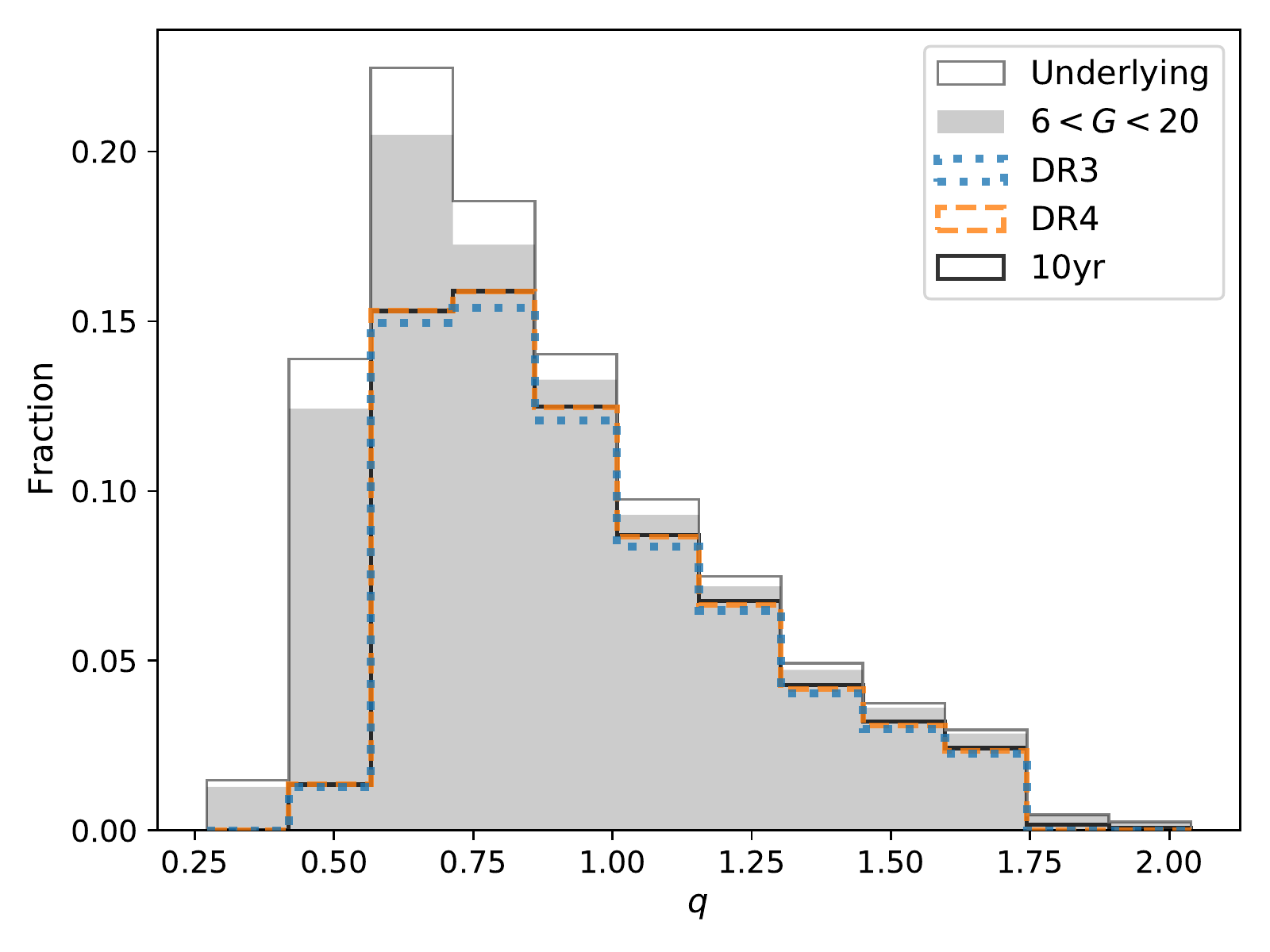}
    \end{subfigure}
    \begin{subfigure}{0.495\linewidth}
    \includegraphics[width = \textwidth]{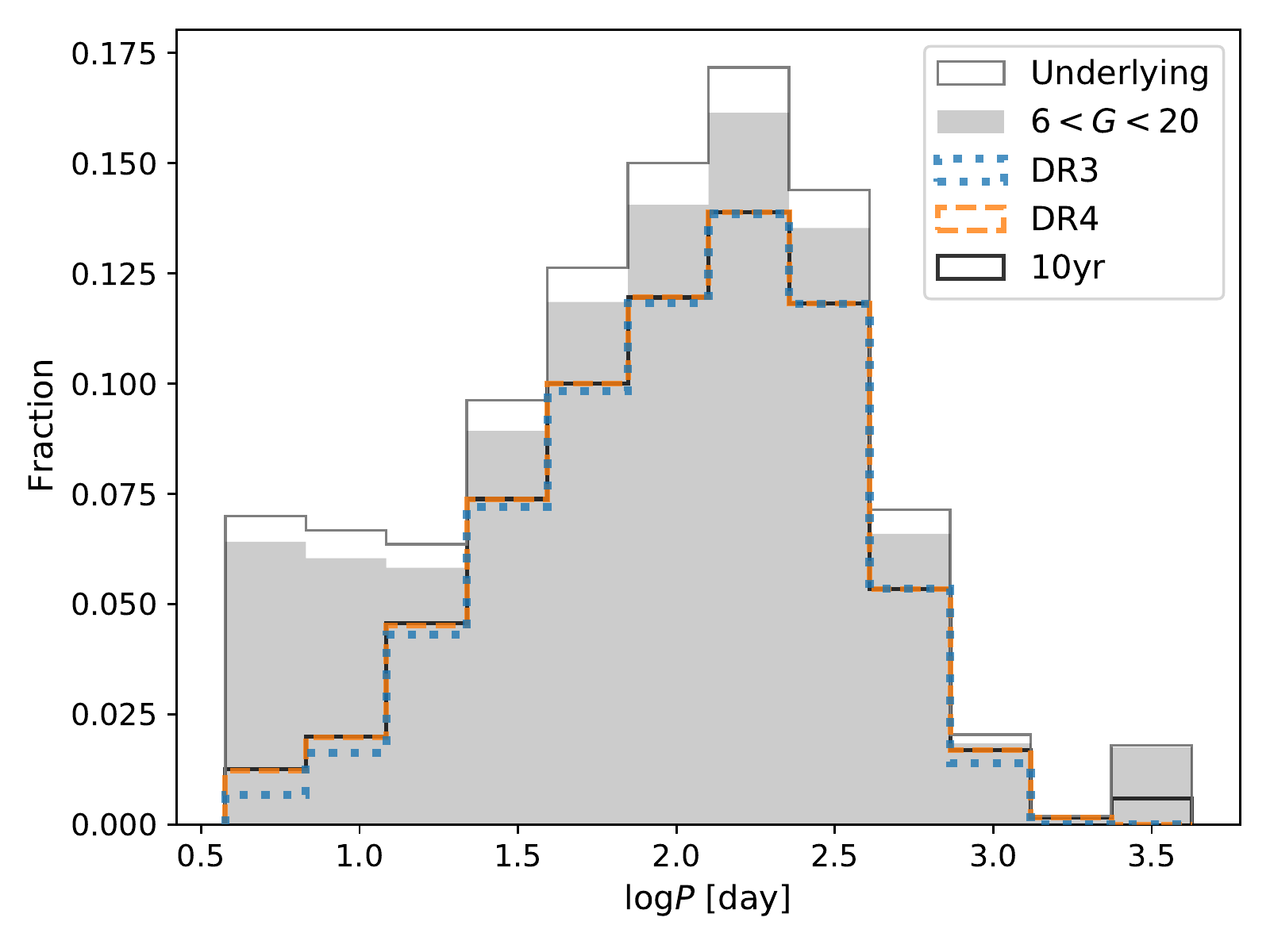}
    \end{subfigure}
    \caption{Same as \Figref{fig_observed_distributions}, but only for the identified OB+BH binaries at the end of the extended 10-yr mission. The distributions of identified systems in the other two data releases are also shown as a reference.}
    \label{fig_observed_distributions_10yr}
\end{figure*}

Figure \ref{fig_observed_distributions_10yr} shows the distributions in masses and periods expected for the identifiable OB+BH systems at the end of the extended 10-yr mission.

\end{appendices}

\end{document}